\documentstyle[supertabular,epsf,rotate,wrapfig,graphicx]{mn}
\input epsf.sty

\newcommand*{\twosim}{\mathrel{\vcenter{\offinterlineskip\hbox{$\sim$}\vskip-.35ex\hbox{$\sim$}}}}

\def\lesssim{\mathrel{\hbox{\rlap{\hbox{\lower2pt\hbox{$\sim$}}}\raise2pt\hbox{$<$}}}}
\def\grtsim{\mathrel{\hbox{\rlap{\hbox{\lower2pt\hbox{$\sim$}}}\raise2pt\hbox{$>$}}}}
\newcommand{\gtsim}{\mbox{{\raisebox{-0.4ex}{$\stackrel{>}{{\scriptstyle\sim}}$}}}}
\newcommand{\ltsim}{\mbox{{\raisebox{-0.4ex}{$\stackrel{<}{{\scriptstyle\sim}}$}}}}

\begin{document}

\title[The 37 brightest radio sources in the SXDF]
{The Subaru/\textit{XMM-Newton} Deep Field - II. 
The 37 brightest radio sources}

\author[Vardoulaki et al.]{
Eleni Vardoulaki$^{1}$, Steve Rawlings$^{1}$, Chris Simpson$^{2}$, 
David G. Bonfield$^{1}$,
\\
\\
 {\LARGE\rm R. J. Ivison$^{3,4}$, Eduardo Ibar$^{4}$} 
\\
\\
$^{1}$ 
Astrophysics, Department of Physics, Denys Wilkinson Building, Keble Road,
Oxford, OX1 3RH, UK \\ 
$^{2}$
Astrophysics Research Institute, Liverpool John Moores
University, Twelve Quays House, Egerton Wharf,\\
Birkenhead CH41 1LD, UK\\
$^{3}$ UK Astronomy Technology Centre, Royal Observatory, Blackford Hill,
Edinburgh EH9 3HJ\\
$^{4}$ Institute for Astronomy, University of Edinburgh, Blackford Hill,
Edinburgh EH9 3HJ\\
}

\maketitle

\begin{abstract}
\noindent
We study the 37 brightest radio sources in the Subaru/\textit{XMM-Newton} 
Deep Field (SXDF). We have spectroscopic redshifts for 24 of 37 objects 
and photometric redshifts for the remainder, yielding a median redshift 
$z_{\rm med}$ for the whole sample of $z_{\rm med} \simeq 1.1$ and a 
median radio luminosity close to the `FRI/FRII' luminosity divide. Using 
mid-IR (Spitzer MIPS 24 $\mu \rm m$) data we expect to trace nuclear accretion 
activity, even if it is obscured at optical wavelengths, unless the obscuring 
column is extreme. Our results suggest that above the FRI/FRII radio 
luminosity break most of the radio sources are associated with objects that 
have excess mid-IR emission, only some of which are broad-line objects, 
although there is one clear low-accretion-rate object with an FRI radio 
structure. For extended steep-spectrum radio sources, the fraction of objects 
with mid-IR excess drops dramatically below the FRI/FRII luminosity break, 
although there exists at least one high-accretion-rate `radio-quiet' QSO. 
We have therefore shown that the strong link between radio luminosity (or 
radio structure) and accretion properties, well known at $z \sim 0.1$, 
persists to $z \sim 1$. Investigation of mid-IR and blue excesses shows that 
they are correlated as predicted by a model in which, when significant 
accretion exists, a torus of dust absorbs $\sim$30\% of the light, and the 
dust above and below the torus scatters $\gtsim$1\% of the light.
\end{abstract}


\section{Introduction}
\label{intro}

There is growing evidence that all massive galaxies contain a supermassive 
black hole (SMBH) at their centres (Ferrarese \& Merritt 2000; Gebhardt et 
al.\ 2000; Magorrian 2006), which implies that these galaxies have all 
undergone at least one episode of AGN activity. It is believed that this AGN 
activity is the observational manifestation of feedback processes regulating 
galaxy and cluster evolution (Best et al.\ 2005). At low redshifts radio jets 
emerge from the vicinities of these black holes, where there are indications 
for a `duty cycle' for active jet activity that varies from 
$\sim 0.1 \rightarrow 0.3$ over the 1.4-GHz luminosity range 
$L_{1.4 \rm GHz} \sim 8 \times 10^{23} \rightarrow 8 \times 10^{21} \rm 
W Hz^{-1} sr^{-1}$ -- a critical range as it provides a significant fraction 
($\sim$25\%) of the luminosity density of the population.

Powerful ($L_{1.4 \rm GHz} > 8 \times 10^{23} \rm W Hz^{-1} sr^{-1}$) radio 
sources are more common at high ($z \gtsim$1) redshifts. They are believed to 
have central SMBHs with uniformly high accretion rates at the highest radio 
luminosities and relatively low duty cycles (e.g. Rawlings \& Saunders 1991). 
Low-luminosity radio jets can, however, be associated with high-accretion-rate 
systems, and these so-called `radio-quiet' quasars appear to have similar 
FRI-like radio structures to the low-accretion-rate counterparts 
of similar radio luminosity objects (e.g. Heywood et al.\ 2007). At low 
redshift, the 
most massive ($\gtsim$ $10^{8} \rm M_{\odot}$) SMBHs typically have very low 
accretion rates with systematically higher average values at $z \gtsim$ 2, the 
so-called `quasar epoch' (Yu \& Tremaine 2002). These observational results 
fit in with theoretical ideas that a `quasar mode' of feedback is prevalent in 
the distant universe, and that a `radio mode' of feedback is dominant at low 
redshift (e.g. Croton et al.\ 2006). 

In the unified  model for AGNs (e.g. Antonucci 1993), the central region is 
surrounded by a dusty torus which absorbs light and re-emits it in the 
infrared. Above and below the plane of the torus, dust scatters light yielding 
a blue excess (di Serego Alighieri et al.\ 1993; Tadhunter et al.\ 1992; 
Cimatti et al.\ 1993; Jannuzi et al.\ 1995). Such mechanisms make it difficult 
to observe objects viewed through the torus directly in the optical, UV and 
soft X-rays. The torus creates anisotropic obscuration of the central regions 
resulting in two different types of observed objects, type 1 that are viewed 
face-on and type 2 that are viewed edge-on. Here we use mid-IR observations to 
search for evidence of accretion in a manner which is far less dependent on 
orientation. Note, however, that there are claims that hot-dust emission is 
mildly anisotropic due to the effects of high optical depths and a toroidal 
geometry (Granato \& Danese 1994).

Through the years, there have been many attempts to classify radio sources 
according to their observed properties: e.g. their structure, energy output 
and environment. Fanaroff \& Riley (1974) classified radio sources according 
to their structure by measuring the ratio of distance between the regions of 
highest surface brightness on opposite sides of the central galaxy or quasar 
to the total extent of the source up to the lowest brightness contour 
$R_{\rm FR}$. This way, for FRI radio sources we get $R_{\rm FR} < 0.5$ which 
makes them edge-darkened, while FRIIs have $R_{\rm FR} > 0.5$ meaning they are 
edge-brightened. Furthermore, they found a strong correlation of structure 
with radio luminosity: above 
$\log_{10}( L_{178{\rm MHz}}/{\rm W Hz}^{-1} {\rm sr}^{-1}) \twosim$ 25 
($H_{0}=50~ {\rm km~s^{-1}Mpc^{-1}}$, $\Omega_{\rm M}=1$ and 
$\Omega_{\Lambda}=0$)\footnote{This corresponds to 
$\log_{10}( L_{178{\rm MHz}}/{\rm W Hz}^{-1} {\rm sr}^{-1}) \twosim$ 25 in 
our adopted cosmology ($H_{0}=70~ {\rm km~s^{-1}Mpc^{-1}}$, 
$\Omega_{\rm M}=0.3$ and $\Omega_{\Lambda}=0.3$) at the typical redshift 
($z \simeq$ 0.1) of the FRI/FRII division from the Fanaroff \& Riley (1974) 
study.} 
lie the FRIIs, and below that the FRIs. 

Ledlow \& Owen (1996) found that both FRIs and FRIIs live in similar 
environments, but the properties of their host galaxies may influence the 
structural appearance of the radio sources, at least for those near the 
division of the two FR classes. A more refined division of radio sources by 
structure is the one appearing in Owen \& Laing (1989). In this case we have 
Classical Double (CD), Twin Jet (TJ) and Fat Double (FD) radio sources: CDs 
correspond to FRIIs, TJs to FRIs and FDs to FRI/FRII division objects. Study 
of the 3CRR, 6CE, 7CRS and TOOT radio samples at $z \sim 0.5$ (McLure et al.\ 
2004) is consistent with the low-redshift result of Owen \& Laing (1989) that 
FRI (FD \& TJ) sources reside in hosts which are on average $\simeq 0.5$ mag 
brighter than those of classical double/FRII sources of comparable radio 
luminosity.

Other classification schemes, such as the `dual-population' model (e.g. 
Jackson \& Wall 2001), divide objects according to nuclear accretion rate 
following 
the pioneering study of Hine \& Longair (1979). The less radio-luminous 
population is composed of FRIs and FRIIs with weak or absent narrow emission 
lines, and the more radio-luminous population of strong narrow-emission-line 
FRII radio galaxies and broad-line quasars, where the division is at 
$\log_{10}( L_{178{\rm MHz}}/{\rm W Hz}^{-1} {\rm sr}^{-1}) \twosim$ 26 
($H_{0}=70~ {\rm km~s^{-1}Mpc^{-1}}$, 
$\Omega_{\rm M}=0.3$ and $\Omega_{\Lambda}=0.7$) and corresponds to the Radio 
Luminosity Function (RLF) break (e.g. Dunlop \& Peacock 1990). Best et al. 
(2003) investigated the dual-population scheme for radio sources using the 
CENSORS sample, selected from the 1.4 GHz NVSS survey (Condon et al.\ 1998). 

Ogle et al.\ (2006) studied a sample of FRII narrow-line radio 
galaxies from the 3CRR survey at redshifts $z < 1$, in order to investigate 
whether they host hidden quasar nuclei. They 
found that in most of these radio galaxies, absorption in the mid-IR is 
present due to dust from a molecular torus. Almost half of their sample of 
narrow-line FR II radio galaxies is weak at 15 or 30 $\mu \rm m$, 
contrary to single-population unification schemes. They conclude that mid-IR 
weak radio galaxies may constitute a separate population of jet-dominated 
sources with low accretion power. In Vardoulaki et al.\ (2006) 
we expressed the need to study radio sources in the mid-IR in order to 
investigate the existence of any hidden accretion activity in low-luminosity 
radio sources. Spitzer observations at 24 $\mu \rm m$ are ideal for this task, 
since they trace warm dust emission which can be obscured only by extreme 
columns (Ogle et al.\ 2006). 

To address this important issue, we have looked at a complete sample of radio 
sources in the Subaru/\textit{XMM-Newton} Deep Field (SXDF). The sample 
studied here is the 37 brightest radio sources from the VLA survey (SXDS) of 
the SXDF (Simpson et al.\ 2006) with flux densities greater than 2 mJy at 1.4 
GHz. In Section 2 we present new data taken on the SXDS radio sources, namely 
optical and near-infrared imaging and spectroscopy. We also present a 
cross-correlation with the radio sample of Tasse et al.\ (2006), who observed 
this sky region at 325 MHz, as well as GMRT observations at 610 MHz 
(Sec. 2.1). Section 3 presents the analysis of our sample including 
$K$-band/radio overlays (Sec. 3.1), photometric redshift estimation (Sec. 
3.2), and the classification of the SXDS-based 
sample according to optical and infrared data (Sec. 3.3). In Section 4 we 
discuss accretion indicators, based on our optical/infrared radio-source 
classification. In Section 5 we present the conclusions of our study. 
Finally, in the Appendix, we describe properties of each radio source 
individually.

We adopt a radio spectral index $\alpha$ = 0.8 
($S_{\nu} \propto \nu^{-\alpha}$), unless a spectral index could be calculated 
using the 1.4 GHz, 610 MHz and 325 MHz data. We assume throughout a 
low-density, $\Lambda$-dominated Universe 
in which $H_{0}=70~ {\rm km~s^{-1}Mpc^{-1}}$, $\Omega_{\rm M}=0.3$ and 
$\Omega_{\Lambda}=0.7$.

\addtocounter{figure}{0}

\begin{figure*}
\begin{center}
\setlength{\unitlength}{1mm}
\begin{picture}(150,120)
\put(195,-2){\includegraphics{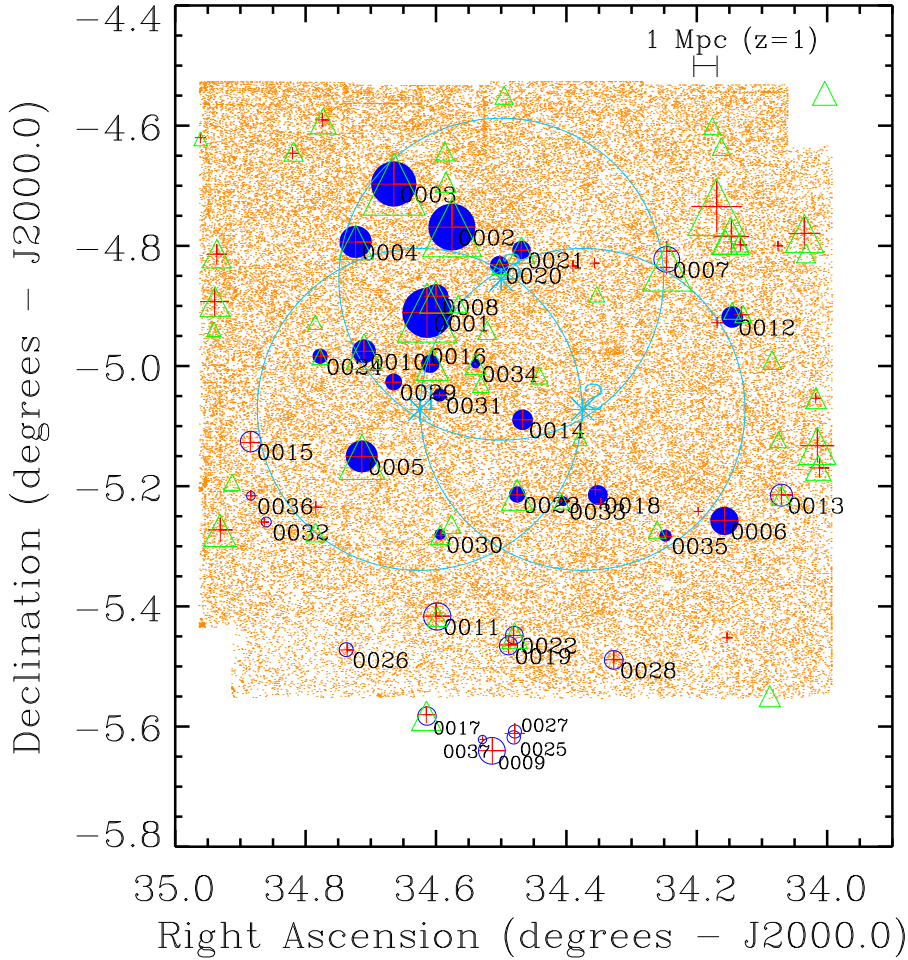}}
\end{picture}
\end{center}
{\caption[junk]{\label{radec} Sky positions of the 37 brightest radio 
sources in the SXDS (blue circles); these are scaled by the logarithm of the 
integrated flux density at 1.4 GHz from the VLA observations of Simpson et 
al.\ (2006) starting at $S_{1.4 \rm GHz}$ = 2 mJy. The shaded area shows the 
UKIDSS-UDS (Lawrence et al.\ 2007) coverage. Note that sxds\_0009, sxds\_0017, 
sxds\_0025, sxds\_0027 and sxds\_0037 lie outside the UDS coverage. Light-blue 
stars are the VLA A-array pointing positions and big light-blue circles 
the FWHM points of the primary beam of the VLA primary beam; objects inside 
these regions have reasonable quality A-array maps and are highlighted with 
filled blue circles. Green triangles are the 325 MHz sources 
($S_{325 \rm MHz} \gtsim$ 4 mJy) in this area (Tasse et al. 2006). Red crosses 
are NVSS (Condon et al.\ 1998) sources in the same sky area, again with the 
symbol size scaled by the logarithm of the NVSS flux density starting at 2 
mJy. The 1 Mpc bar shows the proper physical size of a 1 Mpc-long ruler at 
$z = 1$.
}}
\end{figure*}

\normalsize

\section{Data and Observations}
\label{data}

Optical and X-ray observations of the SXDF (Fig.~\ref{radec}) were made within 
the 1.3 square degree Subaru/\textit{XMM-Newton} Deep Field with Subaru and 
\textit{XMM-Newton} respectively. A follow-up deep radio imaging survey 
at 1.4 GHz was conducted with the Very Large Array (VLA) in B and C 
configurations (Simpson et al.\ 2006). A-array observations are also 
available (Ivison et al.\ 2007). The total radio catalogue, hereafter the 
SXDS, covers 0.8 $\rm deg^{2}$ to a peak flux density limit of 100 $\mu$Jy at 
1.4 GHz. 

The Subaru--XMM-Newton Deep Field (SXDF) has excellent multi-wavelength 
coverage, having been observed in X-rays with the XMM-Newton satellite (Ueda 
et al., in prep.), in the optical $BVRi'z'$ bands with Suprime-Cam on the 
Subaru telescope (Sekiguchi 2004), in the near infrared J and K 
bands as part of the UKIRT Infrared Deep Sky Survey Ultra-Deep Survey 
(UKIDSS - UDS; Lawrence et al.\ 2007), in all of the Spitzer IRAC (3.6, 4.5, 
5.8 and 8.0 $\mu$m) and MIPS (24, 70 and 160 $\mu$m) bands as part of the 
Spitzer Wide-area Infra-Red Extragalactic (SWIRE) survey Lonsdale et al.\ 
(2003) and at sub-mm wavelengths (450 $\mu$m and 850 $\mu$m; Mortier et al.\ 
2006 \& Aretxaga et al.\ 2007). The area is large (for such a deep survey), at 
approximately 1 square degree in the optical and near-infrared, although the 
coverage varies between the different wavebands. The UDS covers approximately 
0.8 $\rm deg^{2}$ centred on the SXDF. 

We study here the brightest 37 radio sources from the 1.4 GHz SXDS 
(Simpson et al.\ 2006); these are named, hereafter, sxds\_0001 to 
sxds\_0037 in decreasing order of 1.4-GHz flux density. Table 1 gives basic 
information about our sample, like their radio position, flux density, 
redshift, optical and radio classification, spectral index, angular size and 
$K$ (2.2 $\mu \rm m$) magnitude (in the AB magnitude system). We adopt the 
radio structure classification from Owen \& Laing (1989), using the VLA data 
to classify objects as CD, TJ, FD or COM (compact) radio sources. 
Wherever available, we used the A-array data to determine whether an object 
has compact hotspots and is therefore a `CD': in the few cases where 
this classification is uncertain, because of no available A-array data, we 
took our best guess based on the appearance of the B-array map and marked 
the classification with a `?'. 
Details of the optical classification are presented in Section 3.3.

In Table 2 we give the result of optical spectroscopy of the SXDS radio 
sources. We have $\approx$ 65\% (24 out of 37) completeness in spectroscopic 
redshifts, with the spectra shown in Fig.~\ref{seds}. Table 2 summarises 
spectroscopic observations carried out on the SXDS objects using either the 
ISIS spectrograph on the WHT telescope, FOCAS on the SUBARU, 2dF or AAOmega on 
the AAT, VIMOS on the ESO-VLT, SITe1 on the Magellan 2 or DEIMOS on the Keck 
2. Calculation of redshifts from VIMOS spectra were done by using the 
{\it fxcor} cross-correlation task in IRAF with SDSS galaxy templates 
(Simpson et al. in prep), apart from sxds\_0025 where the cross-correlation 
failed (see Appendix A). FOCAS spectra have been corrected for atmospheric 
absorption. The object sxds\_0016 has spectroscopy obtained with the DEep 
Imaging Multi-Object Spectrograph (DEIMOS) on the Keck 2 telescope (van 
Breukelen et al.\ 2007).

Table 3 presents the blueness and mid-IR excess, the radio luminosity at 1.4 
GHz $L_{1.4 \rm GHz}$, the luminosity at 24 $\mu \rm m$ 
$[\lambda L]_{24 \mu \rm m}$ and basic quantities for the SXDS radio sources 
used in the analysis (see Sec. 4.1).

Optical and infrared photometric data were used to create SEDs for our 
objects as shown in Fig.~\ref{seds}. Thirteen of our objects are not as yet 
spectroscopically confirmed, so we use photometric redshifts for our analysis 
(Sec. 3.2). In Figure~\ref{radec} we have shown the sky area occupied by the 
37 radio sources in our sample. We are confident that our sample is fairly 
complete. A NVSS-selected 2-mJy sample in the same area gets 34 of the 37 
SXDS sources, and the other four lie just below the formal NVSS 2.3 mJy flux 
density limit.

In order to calculate spectral indices for our sample we use either the 325 
MHz catalogue of Tasse et al.\ (2006), or the 610 MHz flux densities from 
Ibar et al. (in prep), taking into account the radio-source structure 
from the A/B-array maps. 

Tasse et al.\ (2006) surveyed the XMM-LSS field (Pierre et al.\ 2004) in A- 
and B-array VLA configurations at 325 MHz (see Sec. 6). To 
minimise the effect of Doppler-boosted emission, we would ideally have 
selected our objects at 325 MHz, but this survey was unavailable at the time 
the selection was made. We have, however, subsequently cross-matched the 325 
MHz sample with the whole 1.4 GHz radio catalogue. The resolution of the 
325 MHz data is 6.7 arcsec and it has a median 5$\sigma$ sensitivity of 4 
mJy beam$^{-1}$. 
Figure~\ref{radec} shows the 325-MHz sources that are in a sky area somewhat 
larger than our SXDS sample. From our 37-source sample only 22 match with the 
325 MHz catalogue. For all of the matching objects we calculate radio spectral 
indices using the 1.4 GHz and 325 MHz flux densities. Limits on the spectral 
indices are calculated assuming a $S_{325 \rm MHz} \ltsim$ 4 mJy (5$\sigma$) 
limit for the 325 MHz catalogue, although (see Appendix A) there are a few 
cases where this process is unreliable (e.g. sxds\_0011 where 
a steep spectral index is expected from the TJ radio structure of this 
radio source).

The fact that there are 13 325-MHz objects with reasonably high flux density 
that match with lower 1.4-GHz flux density objects than our 2 mJy limit is 
unsurprising, given that $S_{325 \rm MHz} \gtsim$ 4 mJy corresponds to 
$S_{1.4 \rm GHz} \ltsim$ 2 mJy for $\alpha \sim 0.9$ and such steep-spectrum 
objects exist; their number will be exaggerated by random biases on the 
325-MHz flux densities at the limit of the Tasse et al. (2006) survey.

The 610-MHz observations of the SXDF were made with the Giant
Metre-wave Radio Telescope (GMRT) in two
observing runs: on 2006 February 2--5 and 2006 December 4--9 (details
can be found in Ibar et al. in prep). The region was covered with three
pointings forming an equilateral triangle with 22-arcmin sides,
covering an area of $\sim$0.5 deg$^{2}$. Adding both sets of
observations, the total integration time per pointing was 11.4 hr.

To calibrate the data, we used 3C 147 or 3C 48 as flux and bandpass
calibrators, observed for 20 min at the beginning and end of each
track. We also observed 0116$-$208 or 0240$-$231 for 7 min every
45 min to be used as phase and amplitude calibrators.  The full
available bandwidth of 32 MHz was split into two intermediate
frequencies (IFs) of 128 channels each, and reduced independently.

For the data reduction we used the Astronomical Image Processing
System (AIPS). Standard tasks (e.g.\ {\sc quack, tvflg, uvflg, spflg})
were used to flag bad baselines, antennas, initial integrations, and
channels that suffered from narrow-band interference.  A first rough
calibration using {\sc calib} and {\sc clcal} was performed before
using the {\sc flgit} task to flag more bad data using a noise-based
criterion. The clean data were calibrated again and used to normalise
the target using {\sc getjy}. Each IF was compressed in chunks of 7
channels (to ensure bandwidth smearing was not a problem) leaving 15
averaged channels which were used for imaging.  Each pointing was
broken into 37 facets, in order to minimise 3D non-coplanar issues,
and imaged using {\sc imagr}.

The GMRT has a large number of antennas in the central region. These
dominate the $uv$ coverage and result in a non-Gaussian beam
shape. Baselines shorter than 1.5 k$\lambda$ were omitted from the
imaging. We utilised one self-calibration in phase and two in phase
and amplitude, averaging both polarisations in these two last passes,
in order to obtain the final image facets. These were combined using
{\sc flatn}, after convolving them to a single, common beam size. The
final radio image reaches an r.m.s.\ of $\sim 60$ $\mu \rm Jy$ $\rm beam^{-1}$ 
near the centre of the field, with a synthesised beam of
6.8 $\times$ 5.4 arcsec$^{2}$ at position angle of 30 deg.  The
south-west field, centred at R.A.\ 2h 18m 44.0s, Dec.\ $-5^{o}$
7$'$ 20$''$ has a considerably larger mean r.m.s. ($\sim 100 \mu$Jy)
than the other two fields. Also, the noise in the image increases near
bright sources, where it can be as high as $\sim 300$ $\mu$Jy r.m.s.,
limiting the detection of nearby sources.

\section{Analysis}
\label{analysis}

\subsection{Basic information}
\label{info}

In our analysis we use the radio flux densities from Table 1 according to the 
following criteria. In the cases where the NVSS flux density is higher than 
the SXDS one, and the 
source is not compact, we use the NVSS flux density for our analysis, since 
the source may have more structure than the B-configuration VLA observations 
could reveal. On the other hand, when the NVSS flux density is higher than the 
SXDS flux density and the source is compact we assume that the NVSS 
measurement is inaccurate due to confusion by a nearby source (which we 
confirmed by inspection of the VLA image); in this case we use the 
SXDS flux density. Finally, when the SXDS flux density is higher than the 
NVSS value, and the source has compact radio structure, we assume that the 
source is time variable; in that case we also use the SXDS flux density for 
our analysis.

For every SXDS radio source $K-$band UDS images were downloaded from the WFCAM 
archive (Hambly et al. 2007), with the exception of sxds\_0009, sxds\_0017, 
sxds\_ 0025, sxds\_0027 and sxds\_0037 that were not observed by the UDS 
(Fig.~\ref{radec}). For these objects we use $z-$band images from Subaru in  
order to produce overlays of the radio B-array VLA data and imaging, as shown 
in Fig.~\ref{seds}. Also in Fig.~\ref{seds} we present spectral energy 
distributions (SEDs) of the objects created as explained in Section 3.2.

\begin{table}
\begin{center}
\setlength{\unitlength}{1mm}
{\caption[Table~\ref{table:sample}]{Basic properties of the 37 brightest 
SXDS radio sources; values in bold are those used in the analysis. 
{\bf Columns 1, 2 \& 3} give the name of the object 
and its VLA (1.4 GHz) radio position (J2000.0) from Simpson et al. (2006). 
{\bf Columns 4 \& 5} give the integrated flux densities at 1.4 GHz from 
Simpson et al. (2006) and NVSS respectively; radio sources labelled with `x' 
suffer from confusion by a nearby source, which results in a higher NVSS flux 
density than the SXDS value as seen in objects sxds\_0025 and sxds\_0027. 
Objects sxds\_0034, sxds\_0036 and sxds\_0037 are not in the NVSS catalogue 
since they are fainter than the 2.3 mJy limit of the survey (Condon et al.\ 
1998); their flux densities were measured from the NVSS maps and are labelled 
with `n'. {\bf Column 6} gives 
the 610 MHz flux density in mJy (Ibar et al. in prep); the character `s' shows 
that the flux densities of each component of the radio source have been added 
to calculate the whole flux density of the radio sources; `(nif)' means that 
the object is not inside the field that was observed; values in brackets are 
inaccurate since they miss part of flux density coming from the source. 
In {\bf Column 7} we give the 325-MHz flux density from Tasse et al. (2006); 
values in brackets are not accurate since they miss part of the flux density 
coming from the source. {\bf Column 8} shows our radio 
source counterpart optical/IR classification (see Sec. 3.3): `Q' stands 
for a quasar, `OQ' for an obscured quasar, `G?' for a possible galaxy, 
`G' for a secure galaxy, `SB' for a starburst, `WQ' for a weak quasar 
and `BL' for a BL Lac object. {\bf Column 9} gives the spectroscopic redshift, 
when available: `?' denotes objects with spectroscopy yielding plausible but 
insecure redshifts; the symbol `+' indicates clear broad emission lines and 
`a' denotes a broad-absorption-line QSO (BALQSO). {\bf Columns 10 \& 11} give 
the photometric redshift calculated with HYPERz using Coleman (CWW) and 
Bruzual-Charlot (BC) templates respectively; brackets are used to denote 
unreliable fits. {\bf Column 12} gives the radio spectral index from 1.4 GHz 
to 610 MHz or from 1.4 GHz to 325 MHz, which are marked with `m'; we use the 
1.4 GHz flux densities in bold for the calculations. Estimated values are in 
brackets. {\bf Column 13} gives the largest angular size in units of arcsec 
and {\bf Column 14} gives the radio classification 
(R Cl) of each object (Owen \& Laing 1989): CD stands for classical double; 
TJ for twin-jet; FD for fat double; and COM for a compact radio source. 
}}
\end{center}
\end{table}
\addtocounter{table}{-1}

\clearpage
\newpage

\begin{table*}
\begin{center}
\setlength{\unitlength}{1mm}
{\caption[Table~\ref{table:sample}]{
(continued)
}}
\begin{picture}(150,50)
\put(270,-195){\includegraphics{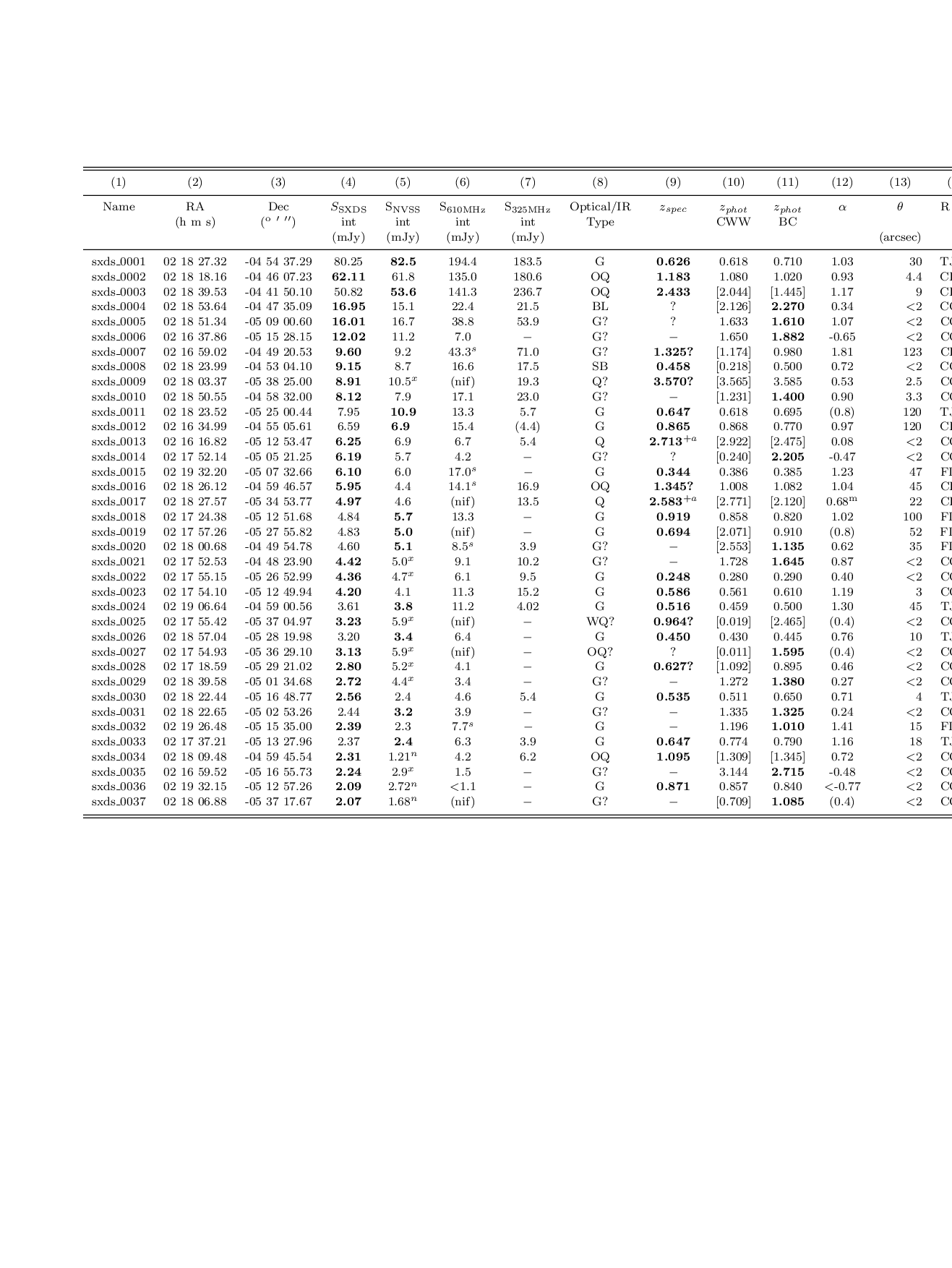}}
\end{picture}
\end{center}

\end{table*}

\normalsize
\addtocounter{table}{0}
\clearpage
\newpage

\scriptsize
\begin{table*}
\begin{center}
{\caption[Table~\ref{spectralog}]{
Observing log for optical spectroscopy of the SXDS radio sources: 
{\bf Column 1} gives the name of the object. {\bf Column 2} gives the 
optical position of the radio sources in RA (h m s) and DEC ($^{\circ}$ m s). 
{\bf Column 3} shows the instrument and telescope where spectroscopic 
observations were 
conducted; either with ISIS at the WHT ($www.ing.ias.es$), with FOCAS at the 
SUBARU ($www.subarutelescope.org$), with VIMOS at the ESO-VLT 
($www.eso.org/instruments/vimos/$), with 2dF/AAOmega at the AAT 
($www.aao.gov.au/AAO/$), with SITe1 at the Magellan 2 
($www.ociw.edu/Magellan$), or with DEIMOS at Keck 2 
($www.keckobservatory.org$). {\bf Column 4} gives the observing date and 
{\bf Column 5} the spectroscopic redshift (a `?' is added when the redshift 
not reliable). {\bf Column 6} gives the 
exposure time; in the case of the WHT we give exposure time for both the (`B') 
and (`R') parts of the spectrum.
}}

\begin{picture}(50,180)
\put(-240,-535){\includegraphics{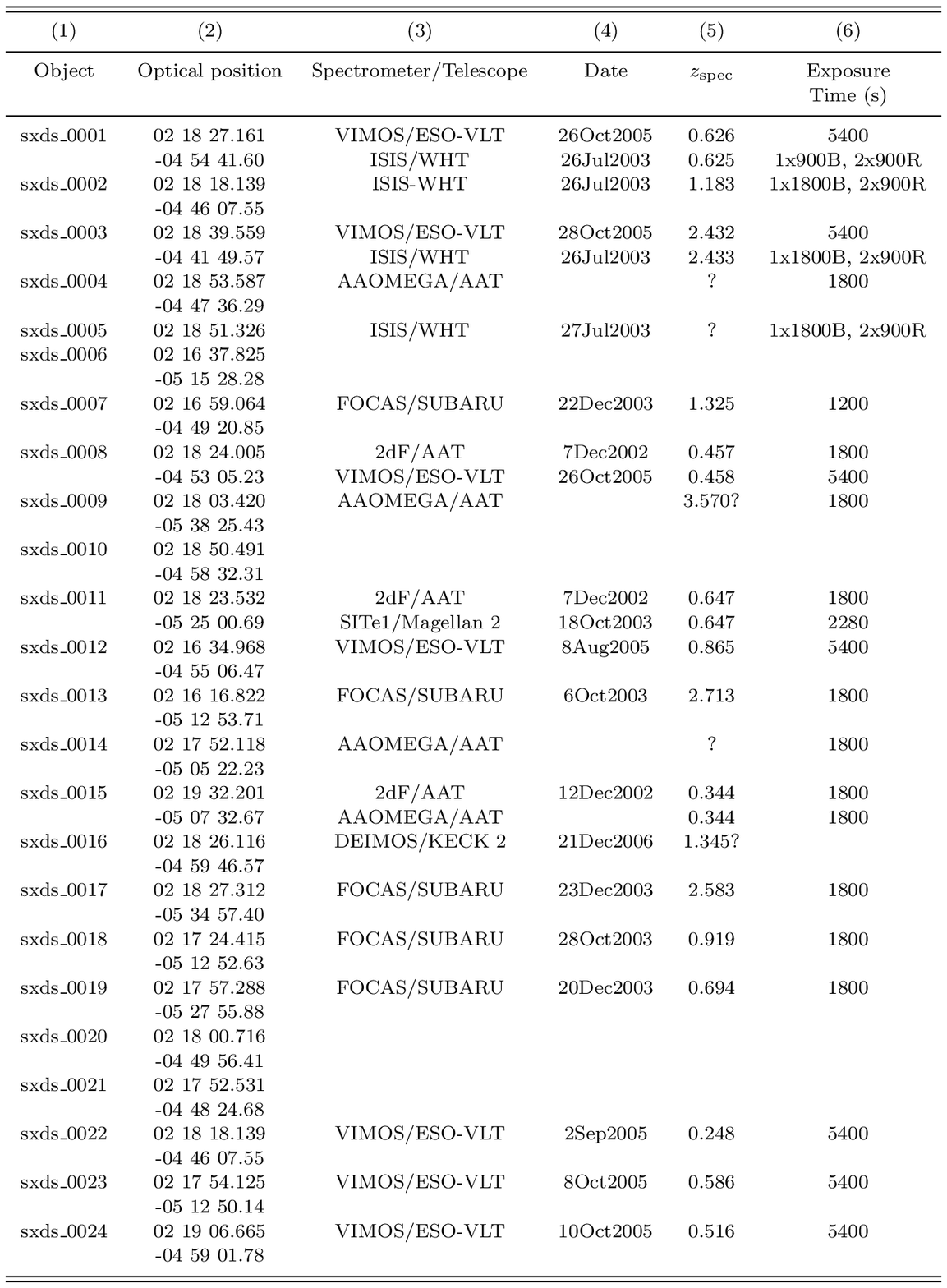}}
\end{picture}
\end{center}

 \end{table*}

\normalsize
\addtocounter{table}{-1}

\clearpage
\newpage

\begin{table*}
\begin{center}
\setlength{\unitlength}{1mm}
{\caption[Table~\ref{spectralog}]{
(continued)
}}
\begin{picture}(50,80)
\put(-80,-180){\includegraphics{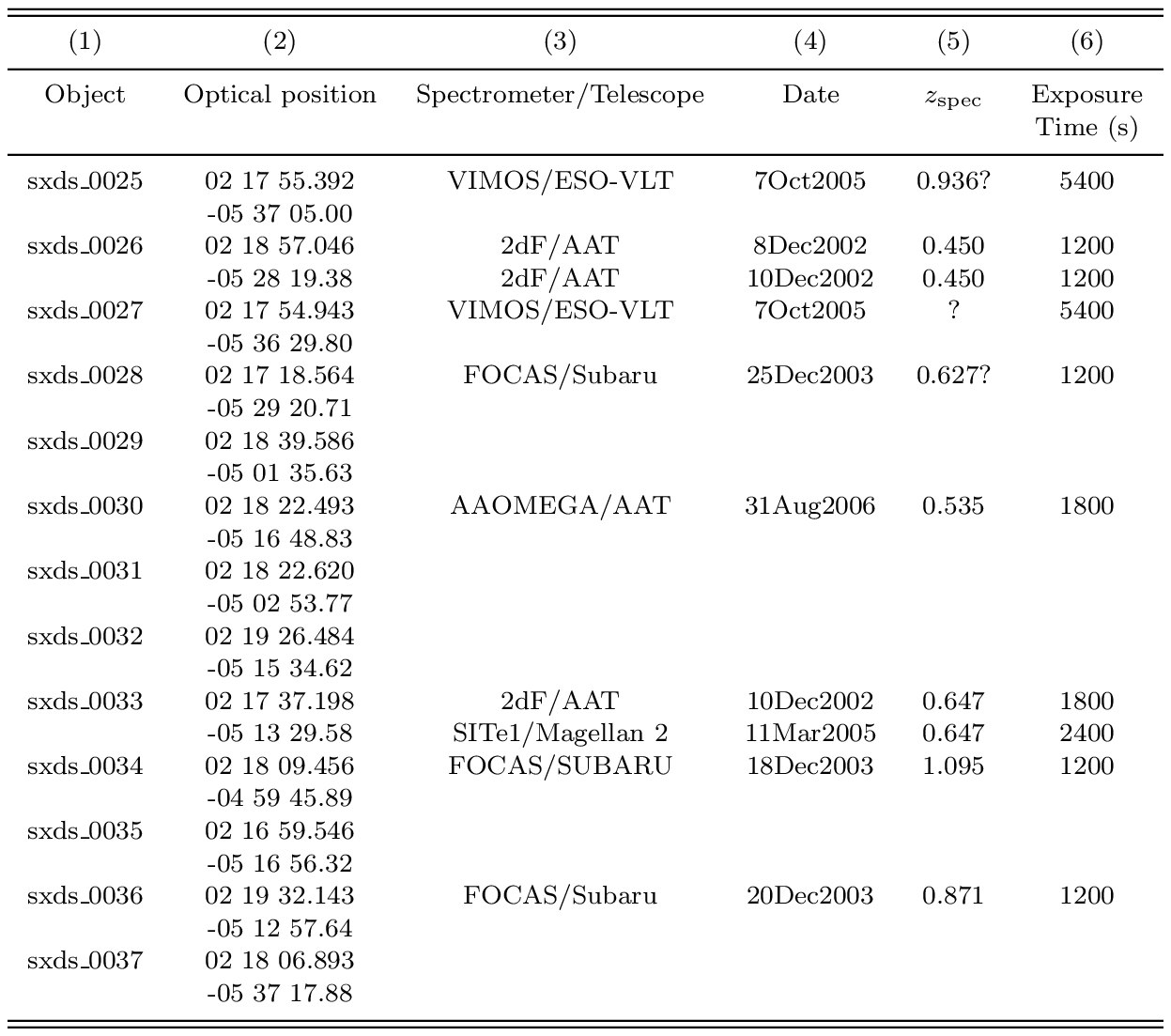}}
\end{picture}
\end{center}

\end{table*}

\normalsize
\addtocounter{table}{0}

\clearpage
\newpage

\begin{table*}
\begin{center}
\setlength{\unitlength}{1mm}
{\caption[Table~\ref{table:sample}]{
{\bf Column 1} gives the name of the object. {\bf Columns 2 \& 3} are the blue 
excess 
($\log_{10}([\nu L]_{4000 \rm \AA rest} / [\nu L]_{1\mu \rm m rest})$) and 
the mid-IR excess 
($\log_{10}([\nu L]_{10 \mu \rm m rest} / [\nu L]_{1\mu \rm m rest})$) 
respectively. {\bf Columns 4 \& 5} give the logarithms of the rest frame 
luminosity at 1.4 GHz and at 24 $\mu \rm m$ respectively. {\bf Column 6} is 
the $K_{AB}$ magnitude from the UDS; the character `$K$' 
denotes point sources at the $K-$band, whereas `o' denotes point sources in 
the optical. For sxds\_0009, sxds\_0017, sxds\_0025, sxds\_0027 and sxds\_0037 
we give an estimate of the $K$ magnitude from the SED of the object (see 
Fig.\ref{radec}); these values are in brackets.
}}
\begin{picture}(150,50)
\put(-30,-240){\includegraphics{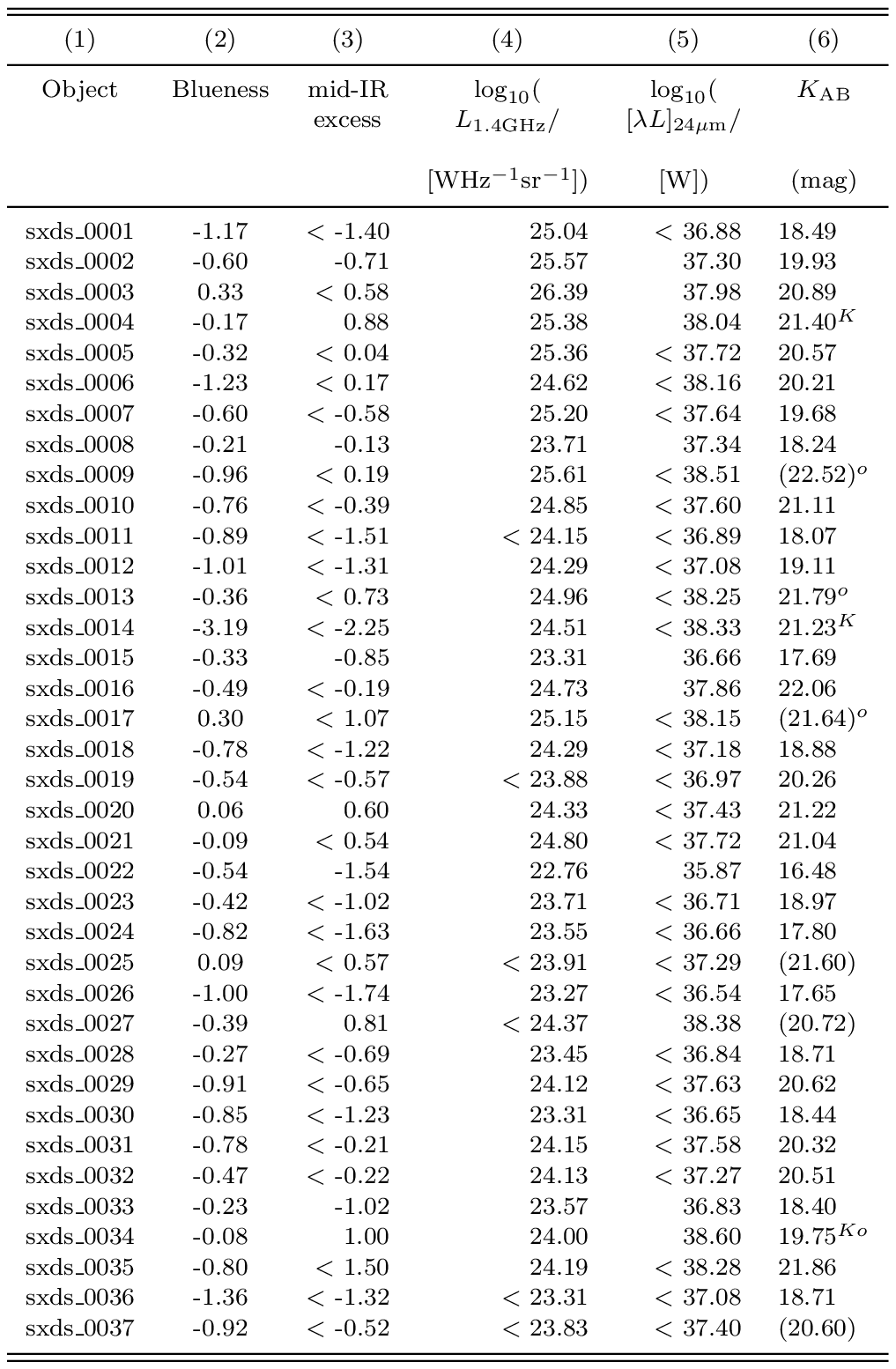}}
\end{picture}
\end{center}

\end{table*}

\normalsize

\clearpage
\oddsidemargin 0.in
\evensidemargin 0.in
\normalsize

\subsection{Photometric redshifts}
\label{photoz}

We have estimated redshifts for the objects by fitting template spectral 
energy distributions - SEDs to their observed photometry, using a modified 
version of the publicly-available\footnote{webast.ast.obs-mip.fr/hyperz/} 
HYPERz code, by Bolzonella et al.\ (2000). HYPERz offers a convenient 
interface for the provision of photometric bandpass functions, galaxy 
templates and data, and is well documented.  It also has built-in 
prescriptions for reddening of templates by dust, and for Lyman forest 
absorption below 912 $\rm \AA$. 

For the purposes of galaxy template fitting, we consider only the optical 
(Subaru), near-infrared (UKIDSS-UDS) and mid-infrared (SWIRE) data. The 
approximate 5-$\sigma$ point source depths achieved in each waveband are 
shown in Table 4.

\addtocounter{table}{0}

\begin{table}
\begin{center}
\caption[Point source depths for SXDF optical, near-infrared and mid-infrared
bands.]{Approximate 5-$\sigma$ point-source depths for the bands used in the
SXDF, and, for the Spitzer bands, the flux density limits of the SWIRE 
catalogues (used as upper limits in the SEDs for undetected objects). }
\label{sxdf-depth}
\begin{tabular}{c|c c|c}
\hline\hline
Band & \multicolumn{2}{|c|}{Depth} & Catalogue flux limit \\
& AB mags & $\mu$Jy & $\mu$Jy \\
\hline
$B$ & 27.5& $3.6\times10^{-2}$& ---\\
$V$ & 26.7& $7.6\times10^{-2}$& ---\\
$R$ & 26.8& $6.9\times10^{-2}$& ---\\
$i'$ & 26.5&$9.1\times10^{-2}$& ---\\
$z'$ & 25.5&0.23& ---\\
$J$ &  23.5&$1.4$& ---\\
$K$ &  23.5&$1.4$& ---\\
3.6 $\mu$m & 22.5 & 3.7 & 10\\
4.5 $\mu$m & 22.1 & 5.4 & 10 \\
5.8 $\mu$m & 19.7 & 48 & 43 \\
8.0 $\mu$m & 20.0 & 37.8 & 40 \\
24 $\mu$m & 18.0 & 230 & 450 \\
\hline\hline
\end{tabular}
\end{center}
\end{table}

The catalogues contain a variety of adaptive and fixed aperture magnitudes, 
which have been measured from the image data using SExtractor (Bertin \& 
Arnouts 1996). For the purpose of SED fitting, it was important to select 
those apertures which provided the most accurate colour information for each 
object. Whilst the adaptive (such as Kron and isophotal) magnitudes provide a 
good estimate of the total flux of an object (without the need for an aperture 
correction to be derived based upon the galaxy's assumed surface brightness 
distribution and the seeing, as is required to correct a small fixed 
aperture), the apertures must generally be large, containing areas of low 
surface 
brightness which increase the photometric error. Also, an adaptive aperture 
might be a different size in each measured band and thus sample a different 
component of the stellar population, potentially leading to unphysical colours.
For these reasons, and because of the problems using adaptive apertures at 
faint magnitudes (where errors in 
determining the correct isophotal or Kron radii can exacerbate photometric 
errors in low signal-to-noise images), we decided to use fixed aperture 
photometry for our SED fitting.

However, fixed apertures have their own issues to consider. In the ideal 
situation, we would have a set of images with the same PSF, perfectly 
registered astrometrically. It would then be possible to extract photometry 
using apertures placed in exactly the same positions in each image, and we 
could be confident that, whatever the aperture size, we would be sampling 
the same fraction of the light from the same regions of each galaxy studied. 
But this is far from the case when combining data taken with different 
instruments, under different seeing conditions, as we are doing here. 
It was undesirable on confusion grounds to degrade the optical and near-IR 
data to match the large PSF of the Spitzer bands, so we chose the aperture 
size to minimise the systematic differences in the fraction of the flux 
measured in each band.  

Clearly, there is a trade-off to be made --- too large an aperture is 
vulnerable to confusion and increased background noise, while an aperture 
which is very small will contain very different fractions of the total flux 
for images with different PSFs.  To investigate this trade off in detail, we 
looked at the mean difference between aperture magnitudes and the 
SExtractor MAG\_AUTO magnitudes, which uses an adaptive Kron aperture 
to estimate the total flux.  For the smallest aperture available in all 
bands (2 arcsec) there were large offsets (of order 0.2--0.3 mags) between 
the aperture and Kron magnitudes, but the most importantly these offsets were 
very different in the different bands. With the next available aperture 
diameter (3 arcsec) the mean differences in the aperture and Kron magnitudes 
were much smaller --- less than a tenth of a magnitude. Deeming this an 
acceptable systematic error, we decided to use 3-arcsec apertures for all 
photometry.  

To combine the catalogues across the optical and infrared bands, we assumed 
that the closest match between extracted objects, within a search radius of 2 
arcsec relative to the radio position (chosen to allow for differences 
in astrometry between different photometric bands), was the correct 
association, with the constraint that a catalogue entry was only permitted to 
match once per band.  

We used both empirical and evolving galaxy SEDs in the template fitting. The 
empirical templates used were the ones supplied with the HYPERz code, which 
are four typical observed galaxy spectra taken from Coleman, Wu \& Weedman 
(1980; henceforth CWW), extended in the ultraviolet and infrared by matched 
models (Bruzual \& Charlot 1993).

The evolving templates were generated using {\sc{Galaxev}} (Bruzual \& Charlot 
2003; henceforth BC\footnote{When running HYPERz with the CWW templates, 
slightly different apertures were used for the magnitudes. An exception is 
sxds\_0016, where magnitudes were measured directly from images and 
then the BC and CWW templates were fitted to the SED.}), with 14 exponentially 
decaying star formation rates 
with e-folding timescales ranging from 0.1 to 30 Gyr, plus an instantaneous 
burst, a finite burst and a constant star formation rate template, all with 
solar metallicity and a Salpeter initial mass function for the stars. This is 
a larger number of BC templates than used by Bolzonella et al.\ (2000), but 
our intention was to span the space of possible galaxy types as finely as 
possible, in the hope that there would always be a template with similar star 
formation history to any object in the dataset, and that this would then be 
the best fit.  

Both empirical and evolving SEDs were used because parallel work by van 
Breukelen et al.\ (2006) seemed to indicate that, while the evolving templates 
gave slightly more accurate results on spectroscopically identified (and thus 
also reasonably bright) samples, they more quickly broke down, giving 
catastrophic redshift estimates, as signal-to-noise was reduced. 
Catastrophically wrong redshift estimates produced by the two sets of 
templates are generally different, so agreement between the results obtained 
with the CWW and BC SEDs is useful to improve confidence that a particular 
redshift estimate is correct. Where the estimates differ significantly, it is 
more often the case that the CWW estimate is closer to reality, though this 
effect is difficult to quantify precisely since catastrophic redshift 
estimates are a function of the template set, as well as the precise 
signal-to-noise in each observed band.  

Figure~\ref{seds} presents the SEDs constructed for our radio sources; red 
colour represents the BC templates and  black the CWW templates. We decided to 
use the BC derived value in the case where no spectroscopy is available.

\begin{figure}
\begin{center}
\setlength{\unitlength}{1mm}
\begin{picture}(150,60)
\put(80,-10){\includegraphics{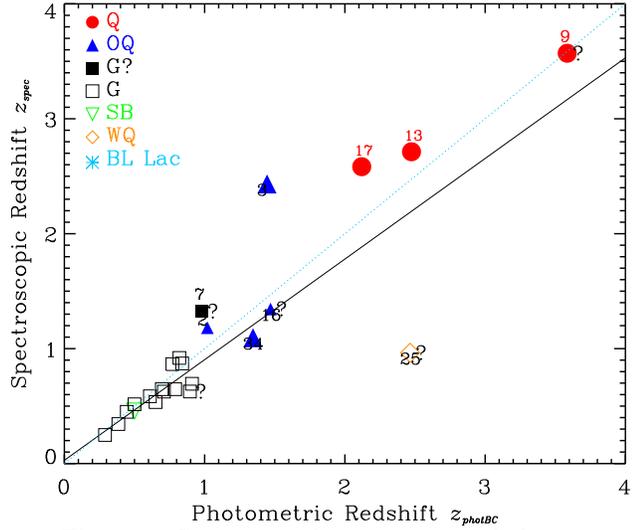}}
\end{picture}
\end{center}
\vspace{0.1in}
{\caption[junk]{\label{zsp-zph}
Comparison of spectroscopic and photometric (based on BC templates) redshift 
in the 24 (of 37) cases where both are available. 
Symbols indicate optical/near-infrared classification as adopted in Section 
3.3, where filled symbols denote objects with spectroscopic redshifts: red 
filled circles for quasars `Q', blue filled triangles for obscured quasars 
`OQ', black filled squares for possible galaxies `G?', black squares for 
secure galaxies `G', green upside-down triangles for starbursts `SB', 
orange diamonds for weak quasars `WQ' and light-blue stars for `BL Lac' 
objects (note that the `BL' object is not 
present due to absence of spectroscopic redshift). The light-blue dotted 
line corresponds to $z_{\rm spec} = z_{\rm photBC}$. The solid black line fits 
the 15 galaxies `G' that have both spectroscopic and photometric redshift: 
$z_{\rm spec-G} = 0.88 \times z_{\rm photBC-G} - 0.03$. Names in red 
represent objects with broad-line optical features. Outlying points are 
labelled with their source number. Objects with uncertain photometric redshifts are plotted with symbols, a factor 1.5 larger than the others.
}}
\end{figure}

In Fig.~\ref{zsp-zph} we compare 24 spectroscopic redshifts with the 
photometric redshifts (Table 2); optical spectra of these objects are 
presented in Fig.~\ref{seds}. Note that objects significantly deviating from the $z_{\rm spec} = z_{\rm phot}$ line have unclear photometric redshifts (see Table 1). Three objects in our sample are classified as quasars (see Sec. 3.3), two of which fall into the $z_{\rm spec} > z_{\rm phot}$ category; the third, sxds\_0009, may have a good photometric redshift because the tentative broad-line emission could be scattered light, with the nucleus making an insignificant contribution at the longer wavelengths dominating the calculation of the photometric redshift. 

The median redshift for our radio-source 
sample, including photometric redshifts, is $z = 1.095$. The best fit line for 
the galaxies `G' gives an excellent agreement between spectroscopic and 
photometric redshifts, within the errors. The linear Pearson correlation 
coefficient is 0.89 and the Spearman correlation coefficient is $\rho = 0.96$ 
with 99.9\% probability for a correlation. This makes us confident 
to use photometric redshifts in the case where no spectroscopy is available. 
Consequently, we adopt the photometric redshifts in bold-type in Table 1. 

\addtocounter{figure}{0}

\begin{figure}
\begin{center}
\setlength{\unitlength}{1mm}
\begin{picture}(140,50)
\put(80,-20){\includegraphics{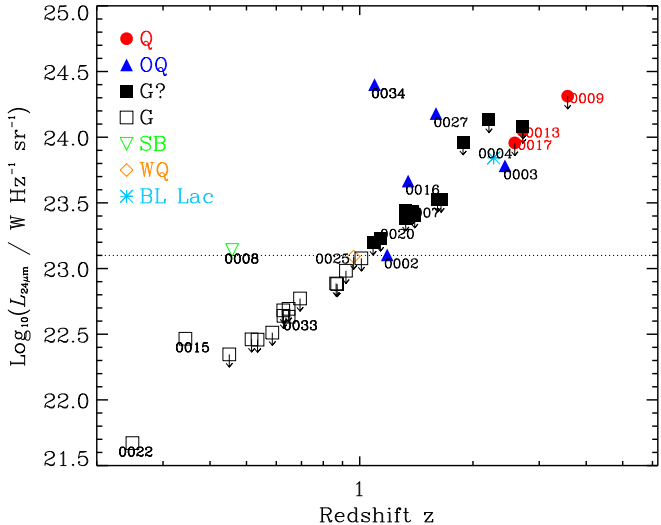}}
\end{picture}
\end{center}
\vspace{0.5in}
{\caption[junk]{\label{l24z_sxds}
Logarithm of the rest-frame Luminosity $L_{24 \mu \rm m}$ at 24 
$\mu \rm m$ versus redshift $z$. Symbols are the same as in 
Fig.~\ref{zsp-zph}. The dotted line corresponds to 
$[\lambda L]_{24 \mu \rm m} = 10^{-1.8} L_{\rm Edd}$ for a quasar with 
$M_{\rm BH}$ $\gtsim$ $10^{8}$ $\rm M_{\odot}$, as discussed in 
Section 3.3. Outlying points are labelled with their source number.
}}
\end{figure}

\addtocounter{figure}{0}

\subsection{Optical/infrared radio-source classification}

Figure~\ref{l24z_sxds} shows the rest frame luminosity of the SXDS radio 
sources at 24 $\mu \rm m$ versus their redshift. The value of 
$\log_{10}( L_{24 \mu \rm m})$ is calculated using the spectral index 
measured from the 24 $\mu \rm m$ data point to the nearest lower-$\lambda$ 
detection in the observed frame SED (Fig.~\ref{seds}).

We use optical/IR observations (Fig.~\ref{l24z_sxds}) to classify a radio 
source as either Quasar `Q', Obscured Quasar `OQ', Galaxy? `G?', Galaxy `G', 
Starburst `SB', Weak Quasar `WQ' or BL Lac `BL', as described below. We deem 
that nuclear accretion is `significant' in objects that obey 
$\log_{10}(L_{24 \mu \rm m}/ \rm [W Hz^{-1} sr^{-1}]) > 23.1$ (or 
$[\lambda L]_{24 \mu \rm m} > 10^{37.3}$ ${\rm W}$)\footnote{Other authors use slightly different values to separate `luminous' quasars (and broad-line radio galaxies) from objects with lower accretion rates (e.g. Ogle et al. 2006 adopt a value of $[\lambda L]_{15 \mu \rm m} \sim 10^{36.9}$ $\rm W$).}. This value corresponds to 
$[\lambda L]_{24 \mu \rm m} \ge 10^{-1.8} L_{Edd}$, a typical lower limit for 
quasars at $z \sim 1$ (see Fig. 2 McLure \& Dunlop 2004), for a black hole 
mass $M_{\rm BH} \ge 10^{8} M_{\odot}$, a typical lower limit for radio 
sources (McLure et al. 2004); $L_{\rm Edd}$ is the Eddington luminosity 
$L_{Edd} = 10^{39.1} \times (M/10^{8}\rm M_{\odot}) W$. 

We then define the following categories:\\
i) {\bf Q}: Broad lines in the optical spectrum (3/37 cases). None 
of these are detected at 24 $\mu \rm m$, although their limits are 
insufficient to rule out significant accretion.\\
ii) {\bf OQ}: Objects with a 24-$\mu \rm m$ detection (5/37 cases) and with 
sufficient $L_{24}$ to represent significant accretion. In all cases we believe that the nucleus seen at 24 $\mu \rm m$ is obscured optically because the spectra show no broad lines. This 
class may be incomplete in that some objects in the `G?' class, as 
described next, have limits above this critical value.\\
iii) {\bf G?}: A galaxy that has a 24-$\mu \rm m$ limit consistent with it 
lying above the 
$\log_{10}(L_{24 \mu \rm m}/$ $\rm [W Hz^{-1} sr^{-1}]) = 23.1$ 
line\footnote{Because of the 24 $\mu \rm m$ flux density limit these objects 
are the high-redshift (and hence, because of the 1.4-GHz flux density limit) 
high-$L_{1.4 \rm GHz}$ sub-set of the objects lacking Spitzer 
24 $\mu \rm m$ detections.} (11/37 cases).\\
iv) {\bf G}: All other objects (15/37 cases) without significant accretion, 
unless they fall into three special categories defined by properties derived 
from considerations of the optical spectroscopy, the SED and the optical 
structure: {\bf SB}: evidence from 
the SED and optical spectra of a starburst component (1/37 cases; see notes in Appendix A); {\bf WQ}: evidence from the 
SED and optical spectra of a quasar component (see notes in Appendix A) but no 24 $\mu \rm m$ detection (1/37 cases); 
{\bf BL}: Featureless red continuum and a point source at $K$ (1/37 cases).\\
Objects labelled as `Q', `OQ' and `G?' include all objects with, or 
potentially with, significant nuclear accretion, and are represented by 
filled symbols in Fig.~\ref{bluered}. Note, however, that the definition of 
`significant' accretion is somewhat arbitrary as illustrated by the case of 
the `WQ' sxds\_0025. Some accretion may be necessary in many, or perhaps all 
FRIs, and especially those near the FRI/FRII divide, if the radio luminosity 
is assumed to be derived from the Blandford-Znajek mechanism (Cao \& Rawlings 
2004).

\section{Discussion}
\label{discuss}

\subsection{Accretion Indicators}

\begin{figure}
\begin{center}
\setlength{\unitlength}{1mm}
\begin{picture}(140,50)
\put(80,-20){\includegraphics{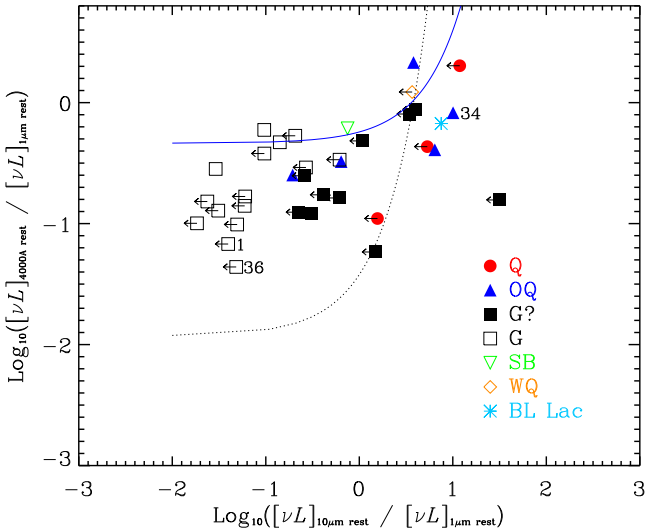}}
\end{picture}
\end{center}
\vspace{0.5in}
{\caption[junk]{\label{bluered}
Blueness versus mid-IR excess for the 37 brightest SXDS radio sources: 
Symbols indicate optical/IR classification, as 
explained in Fig.\ref{zsp-zph} and Sec. 3.3. Table 3 presents measurements 
of those quantities. The black dotted line 
corresponds to the best-fit line in the log-linear plane when all objects 
were treated as detections; the slope and intercept are 0.51 and -1.93 
respectively, giving [see Eqn (2)] $\rm k_{1} \simeq 0.01$ and 
$\rm k_{2} \simeq 0.01$. The blue solid line corresponds to the best-fit line 
in log-linear plane where objects without detections at 24 $\mu \rm m$ were 
treated as limits; the slope and intercept are 0.09 and -0.34 
respectively, giving $\rm k_{1} \simeq 0.46$ and $\rm k_{2} \simeq 0.10$. 
The Buckley-James method in the survival analysis (ASURV) statistics package 
(Lavalley et al. 1992) was used in these calculations. `SB', `BL', `Q' 
and `WQ' objects were excluded from the fits as explained in Sec. 4.1.
}}
\end{figure}

Inspection of the SEDs in Fig.~\ref{seds} shows that some of our objects have 
excess emission at 24 $\mu \rm m$ above that expected from extrapolation of 
the stellar populations. This is quantified via a measure of mid-IR excess, 
$\log_{10}([\nu L]_{10 \mu \rm m rest} / [\nu L]_{1\mu \rm m rest})$ (Table 
3). We chose a rest-frame value of 10 $\mu \rm m$ as our fiducial point as this corresponds to $\sim$ 24 $\mu \rm m$ in the observed frame for redshifts close to the median redshift of our sample. A comparison of mid-IR excess and blueness is presented in 
Fig.~\ref{bluered} where a positive correlation is clear. The generalised 
Spearman correlation calculated using survival analysis statistics ASURV 
(Lavalley et al. 1992) is 0.636 with a 99.9\% probability for a correlation.

Consider a simple model in which blueness is connected to mid-IR excess 
through the following equations:

\begin{equation}
[\nu L]_{4000 \rm \AA \rm rest} = 
\rm k_{1} \times [\nu L]_{1\mu \rm m rest}  + 
\rm k_{2} \times [\nu L]_{10 \mu \rm m rest}   \Rightarrow
\end{equation}
$\log_{10}\left(\frac{[\nu L]_{4000 \rm \AA rest}}{[\nu L]_{1\mu \rm m rest}} \right) =$ 
\begin{equation}
\log_{10}(\rm e) \times \frac{\rm k_{2}}{\rm k_{1}} \times 
\left(\frac{[\nu L]_{10 \mu \rm m rest}}{[\nu L]_{1\mu \rm m rest}}\right) + 
\log_{10}(\rm k_{1})\footnotemark,
\end{equation}
where $\rm k_{1}$ encodes the contribution of the stellar population of a 
passively evolving galaxy formed at high redshift ($z > 5$), and $\rm k_{2}$ 
the mid-IR-excess parameter that we are looking to calculate for this sample 
of radio sources. This model assumes that light from the nucleus with 
intrinsic optical luminosity $L_{\rm opt}$ is i) absorbed by dust and 
re-emitted in the mid-IR generating luminosity $[\nu L]_{10 \mu \rm m rest}$ 
and ii) scattered, generating a contribution to luminosity 
$[\nu L]_{4000 \rm \AA rest}$. \\

\footnotetext{Equation (2) relies on the 
$ln(1+x) \approx x$ approximation which is only 
accurate around and below the knees of the functions plotted in 
Fig.~\ref{bluered}}

\addtocounter{figure}{0}

Objects labelled as `SB', `BL', `Q' and `WQ' have been excluded from the 
fit since 
they have SEDs dominated by different physical processes to those assumed in 
the model described by Eqns (1) and (2). Without spectropolarimetry it is 
difficult to determine whether 
any broad lines are seen as scattered light from the nucleus or they 
represent a direct view of the nucleus. The 24-$\mu \rm m$ luminosity versus 
redshift plane of Fig. 3 shows that all three broad-line objects, as well as 
the `OQ' objects, are consistent with being above or at the 
limit of $L_{24 \mu \rm m} > 10^{23.1}$ ${\rm W Hz^{-1} sr^{-1}}$, or 
$[\lambda L]_{24 \mu \rm m} > 10^{37.3}$ ${\rm W}$.

Fig.~\ref{bluered} shows best-fit lines for two scenarios: 1) all objects were 
treated as detections (black dotted line), and 2) objects are treated as upper 
limits according to their 24 $\mu \rm m$ detection (blue solid line). We note that these `fitted' lines are highly sensitive to the locations in the $\log_{10}([\nu L]_{4000 \rm \AA rest} / [\nu L]_{1\mu \rm m rest})$ vs $\log_{10}([\nu L]_{10 \mu \rm m rest} / [\nu L]_{1\mu \rm m rest})$ plane of a small number of objects, and hence are not robust, although they certainly seem to delineate rough bounds to the data.
 Averaging these results we deduce $\rm k_{1} \sim 0.2$ and 
$\rm k_{2} \sim 0.06$, which agrees well with independent evidence for the 
properties of radio sources as we now argue. The value deduced for 
$\rm k_{1}$ is in line with the expectations of 
template spectra of galaxies which formed their stars at high redshift. 
Optical polarisation studies (e.g. Kishimoto et al. 2001) tell us that 
$[\nu L]_{4000 \rm \AA rest} \gtsim$ $0.01 [\nu L]_{\rm opt}$, 
which is consistent with our value of $\rm k_{2}$ given that QSO SED studies 
suggest $[\nu L]_{10 \mu \rm m rest} \sim 0.3 [\nu L]_{\rm opt}$ 
(Rowan-Robinson 1995), and 0.06 $\times$ 0.3 $\simeq$ 0.01. We conclude that 
whenever nuclear accretion is significant in our sample of radio sources, dust 
in the torus absorbs 30\% of the photons and dust above and below the torus 
scatters $\gtsim$1\% of the photons.

\begin{figure}
\begin{center}
\setlength{\unitlength}{1mm}
\begin{picture}(140,50)
\put(83,-20){\includegraphics{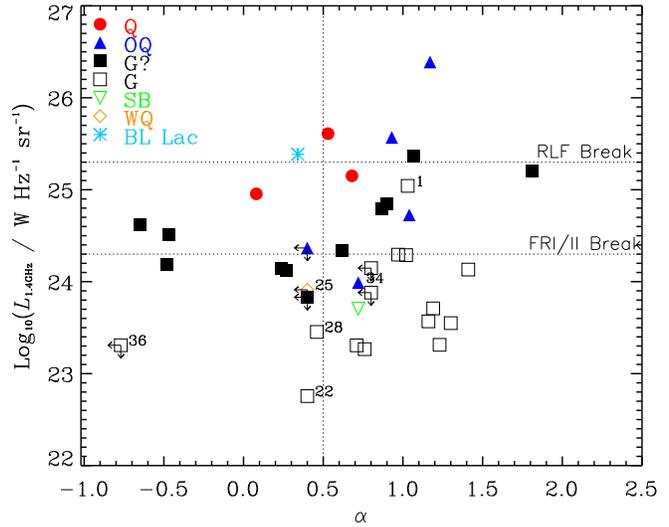}}
\end{picture}
\end{center}
\vspace{0.5in}
{\caption[junk]{\label{la_sxds} Radio Luminosity at 1.4 GHz 
$\log_{10}$ $(L_{1.4 \rm GHz}/$ $\rm [W Hz^{-1} sr^{-1}])$ versus the radio 
spectral index $\alpha$ as in Table 1: symbols are the same as 
in Fig.~\ref{zsp-zph}. The horizontal lines show the RLF and FRI/FRII breaks 
calculated from the Willott et al. (2003) and the Fanaroff \& Riley (1974) 
values respectively using a typical steep-spectrum spectral index of 0.8. The 
vertical line at $\alpha = 0.5$ shows the conventional border between steep 
and flat-spectrum sources. Objects that are discussed in the text (see 
Sec. 4.1) are labelled with their names.
}}
\end{figure}

\addtocounter{figure}{0}

In previous studies, the quasar fraction has been defined as the number of 
sources with quasar-like optical features (e.g. optical broad lines) and has 
a value of $\sim 0.1 \rightarrow 0.4$ over the relevant range of $L_{1.4 \rm GHz}$ 
(e.g. Willott et al.\ 2000). We introduce the term `quasar-mode fraction' 
$f_{\rm QM}$ to describe the fraction of objects with `significant' 
accretion rates, as determined from the observed 24 $\mu \rm m$ flux density, 
to the total number of objects.

In Figure~\ref{la_sxds} we present the radio luminosity at 1.4 GHz of the SXDS 
radio sources against their radio spectral index $\alpha$ (Table 1). A 
typical flat/steep-spectrum radio spectral index $\alpha = 0.4/0.8$, depending 
on whether its radio structure is compact/extended, was used in the case 
where the object was not observed at 610 MHz and is a limit at the 325-MHz 
catalogue. Note that the median radio luminosity is 
$L_{\rm 1.4 GHz} \sim 1.6 \times 10^{24} \rm W Hz^{-1} sr^{-1}$, which lies 
very close to the FRI/FRII luminosity division. Crudely, the SXDS sample is 
sensitive to objects above this division at $z \gtsim 1$ and to objects below 
the FRI/FRII luminosity break at $z \ltsim 1$.

We see from Fig.~\ref{la_sxds} that the single `BL' object  has a flat radio 
spectral index, as expected. It is clear that nearly all the other 
flat-spectrum object have significant accretion, although some exceptions do 
exist, e.g. object sxds\_0028 lies close to the boundary of $\alpha = 0.5$ 
although it has significant errors in the calculation 
of its spectral index. Object sxds\_0022 also lies close to the flat/steep 
$\alpha$ boundary, which is a typical value if star formation is making a 
significant contribution to the radio flux density (see notes in Appendix A). 
Object sxds\_0025 is a `WQ' with a flat radio spectra index and is expected to 
be in that region since we believe we are viewing the central region directly (see notes in Appendix A). The fact 
that it lies below the limit for `significant' accretion is consistent with 
it being a weak quasar. Thus, the only clear exception is sxds\_0036, which is 
a `G' and there is an uncertainty in the calculation of its spectral index. 
The only clear region where the quasar-mode fraction is very low, is for 
steep-spectrum objects below the FRI/FRII break, where $f_{\rm QM} \sim 0.1$. 

\begin{figure}
\begin{center}
\setlength{\unitlength}{1mm}
\begin{picture}(140,50)
\put(83,-20){\includegraphics{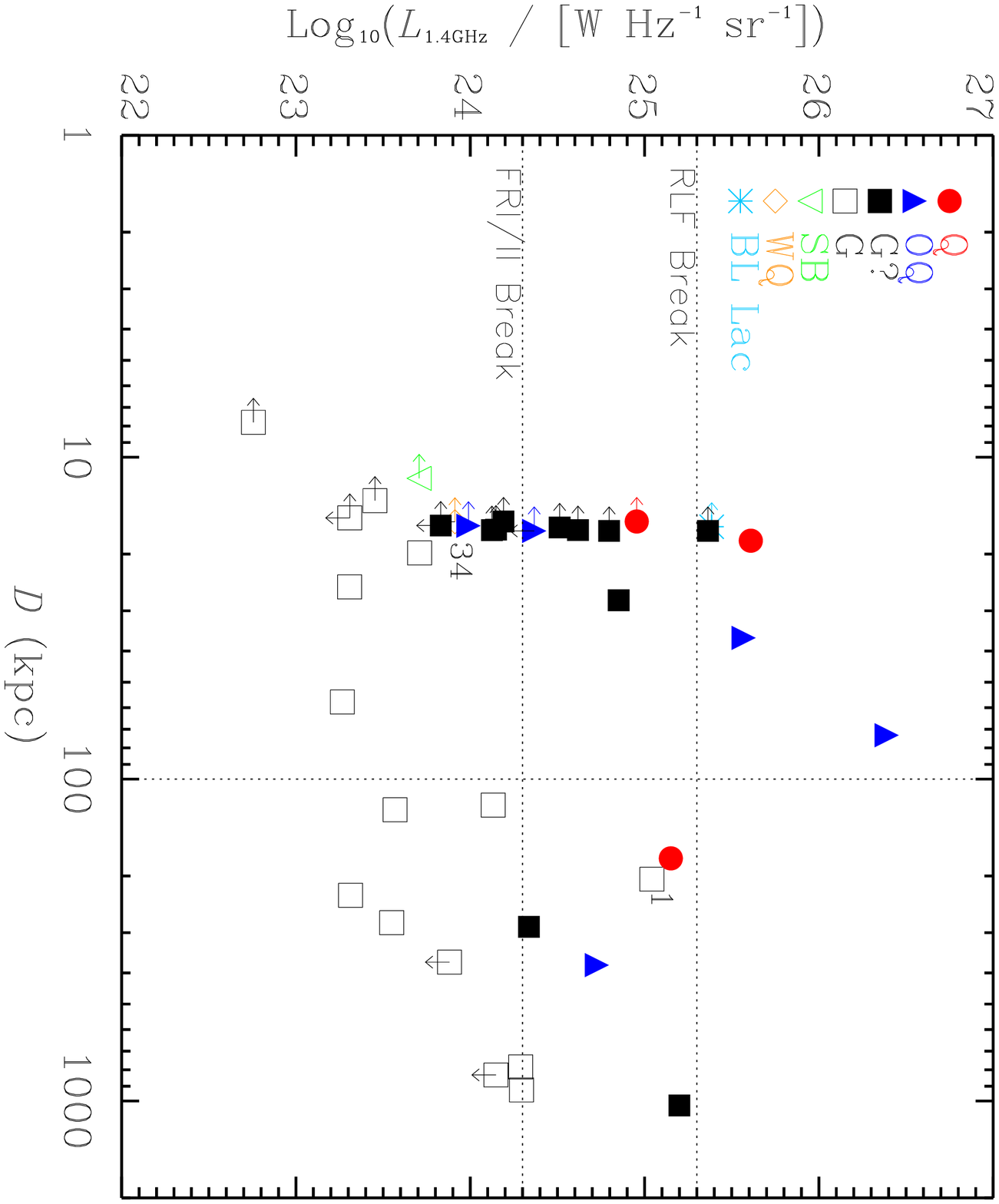}}
\end{picture}
\end{center}
\vspace{0.5in}
{\caption[junk]{\label{ld_sxds} Radio Luminosity at 1.4 GHz 
$\log_{10}$ $(L_{1.4 \rm GHz}/$ $\rm [W Hz^{-1} sr^{-1}])$ versus 
largest projected linear size $D$. Symbols are the same as in 
Fig.~\ref{zsp-zph}. The horizontal lines show the RLF and FRI/FRII breaks 
calculated from the Willott et al.\ (2003) and the Fanaroff \& Riley (1974) 
values respectively using a typical steep-spectrum spectral index of 0.8. 
Objects that are discussed in the text (see Sec. 4.1) are labelled with 
their names.
}}
\end{figure}

Figure~\ref{ld_sxds}\footnotemark \footnotetext{In the similar figure in 
Vardoulaki et al.\ (2006) we used photometric redshifts to estimate values 
of $L_{1.4 \rm GHz}$ used in the plots.}
shows the radio luminosity at 1.4 GHz versus the  projected linear 
size $D$. For SXDS sources that are compact we assume a limiting angular size 
$\theta$ = 2.0 arcsec (Simpson et al.\ 2006).  We see that nearly all `Q', 
`OQ' and `G?' objects of our sample lie above the `FRI/FRII' luminosity 
break with 
the exception of the `OQ' sxds\_0034 (the three `G?' objects near sxds\_0034 
lie close to the boundary of significant accretion).  Above the FRI/FRII break 
we find that $f_{\rm QM} \sim 0.5 - 0.9$ (the lower value assumes the 24 
$\mu \rm m$ 
limits are much higher than the true 24 $\mu \rm m$ values, whereas the higher 
value assumes the true values lie just below the limits). The one clear 
exception in this regime is sxds\_0001, which has no evidence of a QSO and a 
clear Twin-Jet (FRI) radio structure.

\addtocounter{figure}{0}

The quasar-mode fraction drops dramatically below the FRI/FRII break, and 
whether or not one excludes some of the compact ($D <$ 100 kpc) sources as 
potentially part of a separate (beamed) population, then $f_{\rm QM} \ltsim$ 
0.1 because nearly all objects are galaxies `G'. The clear counter-example 
here is sxds\_0034, the only `OQ' below the FRI/FRII break, and potentially 
an optically-obscured example of unobscured FRI QSOs already studied in this 
radio luminosity regime (e.g. Sarazin et al. 1999, Blundell \& Rawlings 
2001). This object is 
spectroscopically confirmed at redshift $z$ = 1.095. An excess of emission 
at 24 $\mu \rm m$ places it securely in the `OQ' classification. This object 
is not a spectroscopically confirmed quasar but has high-excitation narrow 
lines in its optical spectrum and is compact both optically and at $K$; its 
radio map shows a seemingly unresolved source. This seems to be an 
optically-obscured `radio-quiet' quasar in the radio luminosity regime in 
which most radio sources are thought to have FRI radio structures. 
Unobscured `radio-quiet' quasars are known in this regime ( see Heywood et al. 
2007 and references therein), and claims have been made for obscured 
examples (Martinez-Sansigre et al. 2006). However, as found by the study of 
Martinez-Sansigre et al. (2006), the surface brightness sensitivity of 
existing radio data is insufficient to map any extended structure at high 
redshifts. We also note the presence of broad absorption lines in 2 out of our 3 `Q' category objects. This is a high fraction compared to general QSO samples (e.g. Richards et al. 2003 find this fraction to be between $\sim$ 0.03 and $\sim$ 0.2 for optically selected samples of quasars).

Both of Figs~\ref{la_sxds} \&~\ref{ld_sxds} suggest that we have a larger 
fraction of flat, compact radio sources in our sample since it is selected 
at higher frequency than low-frequency selected radio samples like the 
3CRR, 6CE and 7CRS: 51\% $\pm$  7\% objects of our sample have $D <$ 25 kpc, whereas this fraction is $\sim$ 25\% in 7CRS (e.g. Fig. 18 of Blundell, Rawlings \& Willott 1999). The quasar-mode fraction is generally 
high for flat-spectrum objects independent of radio luminosity. Steep-spectrum 
and extended, presumably FRI-type radio sources (i.e. objects below the FRI/II 
break at $\alpha > 0.5$) show no signs of significant accretion with just one 
exception. Cao \& Rawlings (2004) studied a sample of 3CR FRI radio sources 
and found that the $L/L_{\rm Edd}$ ratio in FRI-type radio sources is more 
than two orders of magnitude less than our limit for significant accretion, 
which is in line with our findings.

We have discovered that above the FRI/FRII division the quasar-mode fraction 
is of the order $f_{\rm QM} \sim 1$, where the median redshift of these 
objects is $z_{\rm med} \sim 1.6$. Below the FRI/FRII break 
$f_{\rm QM} \rightarrow$ 0 for extended steep-spectrum sources, 
where $z_{\rm med} \sim 0.65$. A key question is 
whether the presence of significant accretion is linked to the 
radio luminosity $L_{\rm rad}$ (or radio structure) or to the redshift $z$. 
We note that from studies of the 3CR sample (in which the FRI/FRII division 
is at redshift $z \sim$ 0.1 rather that $z \sim$ 1, for our sample) that a 
similar division of mid-IR properties occurs at the same FRI/FRII 
luminosity division (Ogle et al.\ 2006). We conclude that it is a luminosity 
(or structural) effect rather than an epoch effect, and in this paper we have 
shown that such an effect persists from $z \sim 0.1$ to $z \sim 1$. We will 
attempt to de-couple the effects of radio luminosity and radio structure in a 
future paper.

\section{Conclusions}

We have presented a study of the brightest SXDS radio sources with flux 
density at 1.4 GHz above 2 mJy. Our sample is not as yet spectroscopically 
complete (24 out of 37 have spectroscopic redshifts), but Fig.\ref{zsp-zph} 
gives us the confidence in the use of photometric redshifts calculated with 
HYPERz using Bruzual-Charlot templates. The median redshift of the entire 
sample is $z_{\rm med} \approx 1.1$, which from Fig.~\ref{la_sxds} corresponds 
to objects close to the FRI/FRII division. Optical and infrared imaging, as 
well as radio observations are in place to give an insight to the type of 
radio sources at the bright flux density end of the 1.4-GHz SXDS radio survey 
of the SXDF.

We have classified our radio sources as either quasars `Q', obscured quasars 
`OQ', possible galaxies `G?', secure galaxies `G', starbursts `SB', weak 
quasars `WQ' or BL Lac objects `BL'. Objects `Q', `OQ' and `G?' have 
significant nuclear accretion, since they obey 
$\log_{10}(L_{24 \mu \rm m}/ \rm [W Hz^{-1} sr^{-1}]) > 23.1$ (or 
$[\lambda L]_{24 \mu \rm m} > 10^{37.3}$ ${\rm W}$). This value corresponds to 
$[\lambda L]_{24 \mu \rm m} \ge 10^{-1.8} L_{Edd}$ for a black hole mass 
$M_{\rm BH} \ge 10^{8} M_{\odot}$.

The 37 brightest radio sources in the SXDS exhibit the following properties:\\
{\bf $\bullet$} Flat-spectrum, presumably mostly Doppler - boosted, radio 
sources show evidence of accretion in almost all cases, independent of radio 
luminosity.\\
{\bf $\bullet$} Significant accretion is also nearly ubiquitous above the  
`FRI/FRII' luminosity break, i.e. at redshift $z \gtsim 1$ for objects in our 
sample. The quasar-mode fraction, defined here as the fraction of objects with 
significant accretion rates to the total number of objects -- and not as the 
number of sources with quasar-like optical spectra -- is 
$\sim 0.5 \rightarrow 0.9$. There are counter examples like the low-accretion 
FRI sxds\_0001, but they are rare.\\
{\bf $\bullet$} At $z \ltsim 1$ most of the `FRI'-regime sources have low 
accretion rates and $f_{\rm QM} \ltsim$ 0.1, but one high-accretion-rate 
object (sxds\_0034) does exist; this is an obscured `radio-quiet quasar' 
which may have FRI-like jets.\\
{\bf $\bullet$} The presence of significant accretion is linked to the radio 
luminosity $L_{\rm rad}$ (or radio structure) of the radio source at 
$z \sim 1$, just as it is at $z \sim 0.1$.\\
{\bf $\bullet$} Mid-IR excess and blueness are correlated, where the 
Spearman's generalised $\rho =$ 0.636 with a 99.9\% probability for a 
correlation.\\
{\bf $\bullet$} This correlation is consistent with a model in which 
$\sim 30\%$ of the QSO light is absorbed by a dust torus and $\gtsim 1\%$ is 
scattered by dust above and below the torus.\\

Study of the brightest SXDS radio sources in the mid-IR has revealed that the 
ratio of objects with significant accretion rates to the total number of 
objects is high for flat-spectrum radio sources, and for steep-spectrum 
sources above the FRI/FRII radio luminosity divide. This ratio is much lower 
for objects below the luminosity corresponding to the FRI/FRII divide. 
Future work on the brightest SXDS radio sources is likely to involve studies 
of radio structural properties, host galaxies and other indicators of 
accretion like emission lines and X-rays.

\section*{Acknowledgements}
We would like to thank M. Akiyama, O. Almaini, S. Foucaud, H. Furusawa, 
K. Sekiguchi and C. Tasse for practical help with the work described in this 
paper. We would also like to thank the anonymous referee for useful comments. CS would like to acknowledge STFC for an Advanced 
Fellowship. DGB would like to acknowledge PPARC for a studentship.

\appendix

\section{Notes on the 37 brightest SXDS radio sources}
\label{notes}

\oddsidemargin 0.0in
\evensidemargin 0.0in

{\bf sxds\_0001}: The brightest radio source of our sample is a twin-jet (TJ) 
galaxy at redshift $z = 0.627$. Spectroscopy was obtained both with VIMOS on 
the VLT and ISIS on the WHT. In the latter case the slit width was 2.5 arcsec 
and the position angle 
PA = 320$^{\circ}$. This object is resolved at $K$. The NVSS flux density is 
higher than the SXDS value (Table 1), indicating the existence of diffuse 
radio structure surrounding the radio source. This is a typical galaxy `G', 
according to Sec. 3.3, with a prominent [OII] line in its optical spectrum 
and no hint of 24-$\mu \rm m$ emission. The apparent flattening between 610 
MHz and 325 MHz maybe due to inaccurate flux density entry in the 325-MHz 
catalogue.

\begin{figure}
{\caption[junk]{\label{seds}
{\itshape Top:\/} $K$-band images and overlaid radio contours from the 
VLA for sxds\_0001; either the B/C-array (beam $\simeq$ 4 $\rm arcsec^{2}$) or 
the A/B/C-array map (beam $\simeq$ 1.8 $\rm arcsec^{2}$) is shown, with the 
image 
labelled as B or A respectively. For sxds\_0009, sxds\_0017, sxds\_0025, 
sxds\_0027 and sxds\_0037 there are no $K$-band images since these objects 
are beyond the edge of the UDS survey (see Fig.~\ref{radec}), so we use 
$z$-band images instead. The radio structural classification is shown on the 
bottom right of the image. {\itshape Second from top:\/} Overlays of $B-$, 
$R-$ and $z-$band images and B-array radio maps from Simpson et al. (2006). 
{\itshape Second from bottom:\/} Observed frame 
SEDs of our objects, where we also give the best-fit SEDs, their spectroscopic 
redshift $z_{\rm spec}$ (where applicable) and their photometric 
redshift calculated from HYPERz: the red SED is calculated using BC templates 
and the black with CWW templates (see Section 3.2). In all the overlays, North 
is up and East is to the left. The contour levels follow: 
$2^{n_{\rm contour}-1}\times \sigma$, where $\rm n_{contours}$ is the number 
of contours and $\sigma$ the noise level. 
{\itshape Bottom:\/} Optical spectra of 25 of the 37 SXDS radio 
sources. The character `A' denotes atmospheric absorption feature, `S' a 
featured caused by the dichroic, `C' a cosmic ray, and `F' fringing on the 
CCD. All spectra have been smoothed over 3 pixels. The 
following spectra are not flux calibrated: sxds\_0009, sxds\_0028, sxds\_0030.
}}
\setlength{\unitlength}{1mm}
\begin{center}
\begin{picture}(10,10)
\put(10,10){\includegraphics{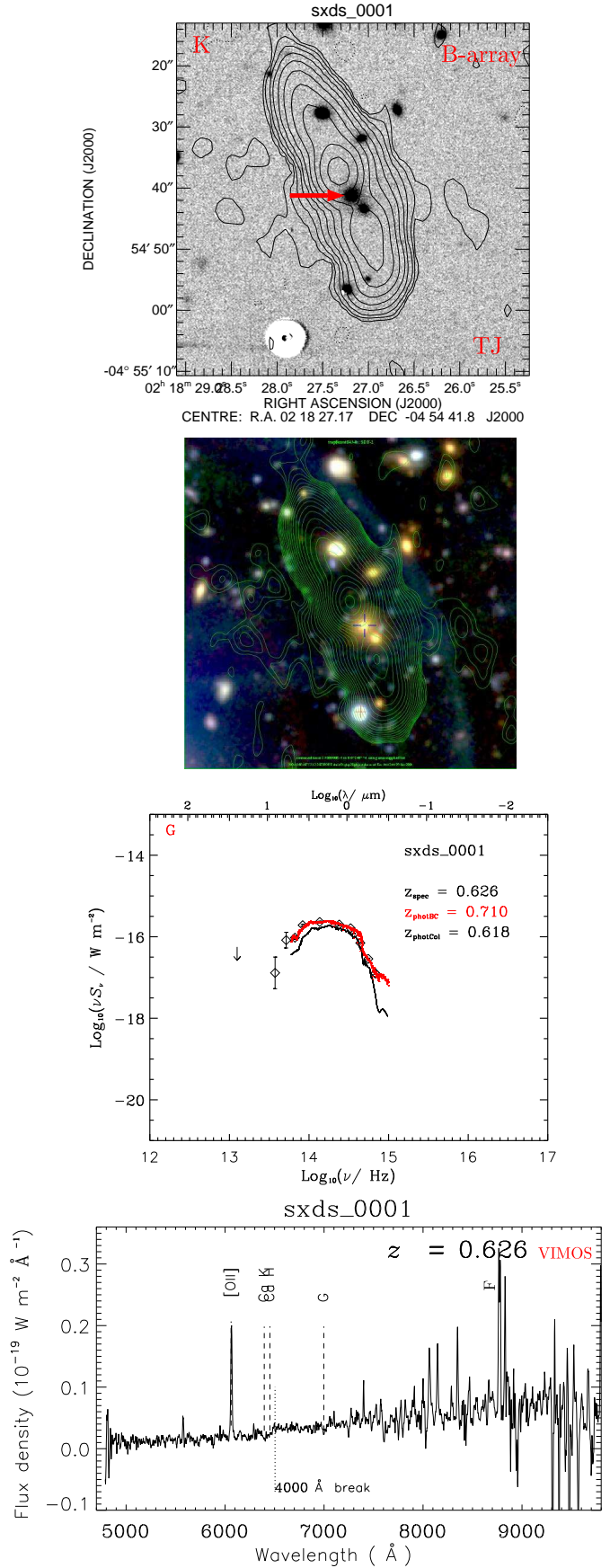}}
\end{picture}
\end{center}

\end{figure}

{\bf sxds\_0002}: This object has a classical double radio structure, as seen 
in the A-array radio map. It is also an obscured quasar `OQ', since it has 
$\log_{10}(L_{24 \mu \rm m}/\rm W) > 37.3$ and is resolved, and presumably 
dominated by starlight at $K$. Optical 
spectroscopy was obtained at the WHT using a slit with a width of 2.5 
arcsec and a PA = 305$^{\circ}$. The BRz-radio overlay in Fig.\ref{seds} 
shows the possibility of merging activity between the radio-source host 
galaxy and a nearby object.

{\bf sxds\_0003}: This is a classical double radio source since the A-array 
radio map reveals compact FRII-like hotspots associated with an obscured 
quasar at redshift 2.433. Spectroscopy was obtained both with VIMOS at the VLT 
and ISIS at the WHT, where in the latter case, 
the slit width was 2.5 arcsec and the PA = 313$^{\circ}$. 
The optical spectrum lacks broad emission lines, a result that confirms 
our classification of this radio sources as `OQ'. As in sxds\_0002, the 
BRz-radio overlay reveals possible galaxy-galaxy merging activity and it has a 
24-$\mu \rm m$ excess in its SED. The Ly$\alpha$ line in the emission 
spectrum of this object is spatially extended over $\sim$ 3.4 arcsec 
$\simeq$ 25 kpc.

{\bf sxds\_0004}: A compact flat-spectrum radio source with a SED rising to 24 
$\mu \rm m$. According to our classification scheme of Sec. 3.3, this is a 
BL Lac object, which is consistent with it having a flat radio spectral index 
($\alpha = 0.164$). The difference between SXDS and NVSS flux densities 
(Table 1) indicates that the radio-source flux density might be time variable. This object 
was observed using AAOMEGA at the AAT, but we only have a 1-D spectrum 
available, with one highly questionable emission line at 5900 $\rm \AA$, 
and this does not give a 
reliable spectroscopic redshift; we prefer thus the photometric redshift.

{\bf sxds\_0005}: A compact steep-spectrum radio source and a galaxy without a 
secure spectroscopic redshift, since observations with the WHT didn't give a 
sufficient S/N to measure an accurate redshift (slit width 2.5 arcsec and 
PA = 310$^{\circ}$). The optical spectrum shows a very faint continuum, as 
expected from its photometry (Simpson et al. 2006) 
and the fact that the photometric redshift, $z$ = 1.610, is in the `redshift 
desert' where there is a lack of any strong emission lines. 

{\bf sxds\_0006}: A compact radio source that is projected close to a bright 
star and 
hence without any entry in the mid-IR Spitzer catalogue. This object 
has no optical spectroscopy, but we are confident that the photometric 
redshift is 
accurate due to a good fit of the template to the SED 
(Fig.~\ref{seds}). It has an inverted spectral index, as calculated from the 
1.4 GHz and 610 MHz data; the limit at 325 MHz verifies this result.

{\bf sxds\_0007}: A large (123 arcsec) FRII classical double radio galaxy at 
$z = 1.325$. The NVSS flux density is the sum of the integrated fluxes 
of the two lobes, which is the same value as the one measured from the 
NVSS map itself. Both 610 MHz and 325 MHz data suggest a very steep radio 
spectral index for this source.

{\bf sxds\_0008}: A compact radio source in a $z = 0.458$ blue galaxy. 
According to our classification of radio sources, this is a starburst `SB' 
with a star formation rate $\rm SFR_{\rm FIR}$ 
$\sim 100$ $\rm M_{\odot} yr^{-1}$, 
calculated from the SED, using a Salpeter IMF and the star formation rate 
(SFR) relations of Kennicutt (1998). We have a good agreement of spectroscopic 
and photometric redshifts. From the BRz-radio overlay of Fig.~\ref{seds} we 
see obvious galaxy-galaxy merging activity and a 24-$\mu \rm m$ excess in 
its SED. We believe the 24 $\mu \rm m$ excess is more readily attributable to a starburst than an AGN because the optical spectrum has bright [OII] emission but no corresponding high-excitation lines.

{\bf sxds\_0009}: A compact radio source missing $J-$ and $K-$band data since 
it lies outside the UDS coverage (Fig.~\ref{radec}). The weak extensions is 
the B-array radio map (Fig.~\ref{seds}) to the east and the south-west are 
probably artifacts. The high NVSS flux density is due to confusion with a 
nearby source. This object is probably at $z = 3.570$, which agrees 
with the photometric redshift (Table 1). The photometry of this object in 
the blue, 
B = 26.3 mag\footnote{($www.astro.livjm.ac.uk/$ $\sim cjs/SXDS/radio/$)}, 
indicates that the apparent blue tail in the optical spectrum is a spurious 
feature. This object has 
a flat radio spectrum and some evidence that a significant fraction of the 
optical light comes from a point source although there are no broad lines or 
a 24-$\mu \rm m$ excess. We classify it as a `Q?' because of hints of broad 
wings to the Ly$\alpha$ line and the significant blue continuum, although the 
limit at 24-$\mu \rm m$ is insufficient to confirm a QSO-like mid-IR excess.

{\bf sxds\_0010}: A compact radio source and a galaxy `G'. The HYPERz 
photometric code gives a good fit to the SED when we use the BC templates. 
There are no spectroscopic data available for this source.

{\bf sxds\_0011}: A radio galaxy at $z = 0.647$ with TJ structure. The angular 
size was measured from the radio overlay from the core to the largest extent 
of the north-east jet and separately to the south-east radio components, 
with the two values 
added together. The 610 MHz flux density was measured around the 
same area as the 1.4 GHz flux density giving a radio spectral index between 
these values of 0.62. The NVSS flux density, though, suggests that there is 
extra emission at large angular sizes. We adopt a typical steep radio spectral 
index of $\alpha$ = 0.8 since steep spectral indices are typical for objects 
with a FRI 
radio structure; the spectral index calculated from the SXDS and 325 MHz data 
suggests an inverted spectral index which is inconsistent with the fact that 
this is a large twin-jet radio source. We suggest that the 325 MHz data 
are inaccurate in this case.

{\bf sxds\_0012}: A large FRII galaxy at $z = 0.865$. The 325-MHz 
catalogue entry includes only part of the source, giving a nonsensical radio 
spectral index. We use the total 610-MHz flux density from the 
radio map to calculate the spectral index, whose steep value agrees with 
the CD radio structure. There are no signs of a QSO nucleus.

{\bf sxds\_0013}:A spectroscopically confirmed quasar at $z = 2.713$ and a 
point source in the optical. The optical spectrum shows strong absorption 
features associated with the broad lines. Such broad-absorption-line (BAL) 
QSOs are rare, but not unknown, in radio-selected samples. The B-array radio 
image shows a compact source. From the optical overlay of Fig.~\ref{seds} it 
looks like a galaxy is dominant in the $K-$band, since it is not a point 
source at $K$, although it is one in the optical. The flat radio spectral 
index, $\alpha = 0.08$, agrees with the fact that this is a compact radio 
source.

{\bf sxds\_0014}: This object appears to be red with a rapid drop in its SED 
from red to blue. In the $K-$band image, although it has a low S/N ratio, 
the source looks unresolved indicating that the radio source might be compact 
or there is a hidden quasar nucleus. The limit at 24 $\mu \rm m$ is 
insufficient to rule out that this might be an obscured quasar `OQ'; we define 
it as a galaxy `G?' (see Sec 3.3). In Vardoulaki, Rawlings \& Simpson 
(2006) we used 36 radio sources from this sample, accidentally excluding this 
one; all the subsequent identification numbers have moved downwards in order 
to include this radio galaxy. The inverted radio spectral index is consistent 
with the fact that this is object has compact radio structure. The tentative 
emission feature around $\lambda \sim 6000 \rm \AA$ in the optical spectrum of 
this object cannot provide us with a reliable spectroscopic redshift. 

{\bf sxds\_0015}: A possible `double-double' radio galaxy. Such sources 
(Schoenmakers et al.\ 1999) are thought to be jets that have re-started, 
sometimes (as perhaps in this case) with a change of jet orientation, 
perhaps due to merging of two supermassive binary black holes (Liu et al.\ 
2003). From the BRz-radio overlay of Fig.~\ref{seds} we see evidence of 
galaxy-galaxy merging in the form of tidal tails (Johnston et al. 2004), and 
there is also excess emission at 24 $\mu \rm m$ as can be seen in its SED. 
We use the radio spectral index calculated between 610 MHz and 1.4 GHz data, 
which gives a steep value that agrees with the FD radio structure. We don't 
trust the 325 MHz data since they give a spectral index inconsistent with 
the radio structure. Although this object is defined as `G', its detection 
at 24 $\mu \rm m$ might indicate it is either a `SB' or a `WQ'.

{\bf sxds\_0016}: A classical-double radio source, possibly at $z = 1.345$. 
Shorter 
wavelengths in the spectrum are, due to poor resolution, affected by the 
overlapping spectrum of a nearby object. Photometry of this object was 
directly measured from images using sxds\_0034 as a boot-strap calibrator, 
since it was not matched with any source within a 2$''$ search radius. Using 
these data we give the photometric redshift in Table 1 with an associated 
uncertainty. According to these measurements, the object has excess emission 
at 24 $\mu \rm m$, which classifies it a `OQ' according to Section 3.3 and 
Fig.~\ref{l24z_sxds}. We use the spectral index calculated between 
the 610 MHz and 1.4 GHz flux densities, which agrees with the CD radio 
structure of this object.

{\bf sxds\_0017}:A spectroscopically-confirmed broad-absorption-line quasar 
at $z = 2.583$, where the Ly$\alpha$ line is heavily absorbed. The spectrum 
appears peculiar, since the CIV and CIII] lines have narrow cores and a red 
and blue broad wing respectively. The radio map reveals a prominent hotspot to 
the north-east; 
the component to the south-west of the quasar is probably the weak side of an 
asymmetric double. We use the 325 MHz flux density to calculate a spectral 
index since it was not observed at 610 MHz.

{\bf sxds\_0018}: A large FD (no compact hotspots in the A-array map) radio 
galaxy at $z = 0.919$. The radio spectral index is calculated between the 610 
MHz and 1.4 GHz data, where the steep value is consistent with the extended 
FD radio structure. The 325-MHz catalogue entry is probably inaccurate, since 
plausible extrapolation of the 610 
MHz flux density suggests that the object has more flux density at 325 MHz 
than the limit of that catalogue.

{\bf sxds\_0019}: A FD (no compact hotspots in the A-array map) radio galaxy 
at $z=0.694$. It has a very asymmetric radio structure. We assume 
$\alpha = 0.8$ due to the extended FD radio structure; the inverted 
spectral index calculated between the 325-MHz and 1.4-GHz data is 
inconsistent with the radio structure of this object.

{\bf sxds\_0020}: This object has an unusual radio structure that we 
classify, uncertainly, as FD. The steep spectral index calculated between 
610 MHz and 1.4 GHz is consistent with the extended radio structure. The 
325-MHz entry is inaccurate since both the radio structure and the 610-MHz to 
1.4-GHz spectral index indicate that it has a steep radio source. There are 
no spectroscopic data available for this object and the photometric estimate 
is highly uncertain.

{\bf sxds\_0021}: A radio galaxy `G' and a compact radio source. The radio 
structure to the north-west side of the object is probably not associated with 
the brightest radio source as it has a separate galaxy identification. The 
photometric redshift is used since there is no spectroscopy available.

{\bf sxds\_0022}: A compact radio source associated with a 
galaxy at $z = 0.248$; note the spectacular blue arc which is probably a 
remnant of star formation associated with a galaxy-galaxy interaction. 
The southern and eastern radio components are 
probably artifacts. This is possibly a post-starburst galaxy, or an `E+A', 
i.e.\ an elliptical with additional absorption lines (e.g. H$\beta$ in Fig.\ref{seds}) in their optical spectra 
due to stars seen $\sim 10^{8}$ yr after the starburst has finished 
(Goto 2007). Star formation might be contributing to the radio emission. 
There is some evidence for excess emission at 24 $\mu \rm m$ as well as 
dramatic tidal tails (Fig.~\ref{radec}). We classify this radio source as a 
`G'.

{\bf sxds\_0023}: A compact radio galaxy at $z = 0.586$. The A-array map 
reveals a possible double structure, although it might be the base of a 
FRI-like jet. There is a fairly good agreement between the spectroscopic and 
photometric redshifts.

{\bf sxds\_0024}: A TJ radio galaxy at $z = 0.516$. The radio spectral index 
that is calculated between 325 MHz and 1.4 GHz is inconsistent with 
the TJ radio structure of the source; so we use 610 MHz flux density.

{\bf sxds\_0025}: A compact weak quasar `WQ' probably at $z = 0.9636$ if the 
single marginally broad line in the spectrum is MgII (the deconvolved 
FWHM is $\sim$ 900 $\rm km s^{-1}$). We are not certain 
that the [OII]$\lambda$2470$\rm \AA$ emission line is real, since 
spectroscopic studies show that this is quite a weak feature (e.g. Serjeant 
et al. 1998). The photometric 
redshift is inaccurate due to the absence of $J-$ and $K-$band data 
(Fig.~\ref{radec}) and the fact that the optical continuum appears to be 
dominated by blue QSO light; the rest-frame EW of the putative MgII line is $\sim$ 50 $\rm \AA$ and thus within the spread seen in directly viewed quasar nuclei (Francis et al. 1999). The templates used in the HYPERz 
code cannot provide a proper fit to the SED. The NVSS flux density is higher 
than expected due to confusion by sxds\_0027. This object was not observed at 
610 MHz since it is outside the coverage of the survey. We assign a typical 
flat radio spectral index $\alpha$ = 0.4 consistent with the compact radio 
structure; we do not trust the 325-MHz catalogue to be complete.

{\bf sxds\_0026}: A TJ radio galaxy at $z = 0.450$. The companion galaxy to 
the west is probably not producing radio emission. According to the radio 
spectral index calculated between 610 MHz and 1.4 GHz data, this is a steep-spectrum 
radio source. The fact that the 325-MHz entry is a limit suggest that we 
should not trust that the 325-MHz catalogue is complete.

{\bf sxds\_0027}: We classify this object as `OQ' since it has excess emission 
at 24 $\mu \rm m$. We use the photometric redshift since the optical spectrum 
shows a featureless continuum. 
The longer-wavelength SED looks more like a BL Lac 
since it can easily be fitted by a power-law with very steep spectral 
index $\alpha \sim 2$, and is consistent with the flat continuum in the 
optical spectrum of this object. The `tails' in the radio map are probably 
artifacts, so its compact nature is consistent either with it being a BL 
Lac object or an `OQ'. The object is resolved at $K$ which is why we 
prefer the `OQ' category. This radio source is not observed at 610 MHz. We 
assign a typical flat radio spectral index $\alpha$ = 0.4 consistent with the 
compact radio structure; we do not trust the 325-MHz catalogue to be complete 
at the flux density level appropriate to this source. 
The NVSS flux density is higher than expected since with the 
NVSS resolution we can only marginally distinguish this object from 
the nearby object sxds\_0025.

{\bf sxds\_0028}: A compact radio galaxy, probably at $z = 0.627$. The radio 
map shows a seemingly double source, but in reality there are two radio 
sources, as there are good optical IDs under both apparent radio lobes. The 
object under the north-west `lobe' is sxds\_0028, whereas the one 
under the south-east `lobe' is sxds\_0042 (a compact radio source at redshift 
$z=0.382$, which is just below our flux density limit). The flattening of 
the radio spectral index between 610 MHz and 325 MHz suggests that the 
325-MHz catalogue is incomplete; the 325 MHz flux should be higher than the 
limit of that catalogue.

{\bf sxds\_0029} A compact radio source that looks unresolved in the $K-$band. 
Due to absence of spectroscopic data, we use the photometric redshift, which 
suggests a fairly good fit to the SED (Fig.~\ref{seds}).

{\bf sxds\_0030}: This probably is a TJ radio galaxy at redshift $z = 0.535$ 
with the jets swept back forming a narrow-angle tail. There is a fairly good 
agreement between the photometric and spectroscopic redshifts.

{\bf sxds\_0031}: This is a compact radio galaxy. The radio component to the 
west is probably not associated with this source as it has its own separate 
galaxy identification. The fairly good fit of the photometric redshift makes 
us confident to use it in the absence of spectroscopic data.

{\bf sxds\_0032}: This is probably a FD radio galaxy; no A-array data 
available to confirm lack of compact hotspots. Without any 
spectroscopic redshift, the BC templates provide a very good fit to the 
photometric data and a believable photometric redshift. The 325-MHz catalogue 
misses this object. According to the spectral index calculated between 
the 610 MHz and 1.4 GHz data, this object should have a flux density at 325 
MHz that would give a steep spectral index.

{\bf sxds\_0033}: A TJ radio source at $z = 0.647$. It belongs to a small 
group of galaxies that is interacting and probably merging with a larger group 
of galaxies (Geach et al.\ 2007). The feature at 
5426 $\rm \AA$ in the optical spectrum (Fig.~\ref{seds}) is CIII] emission 
from a lensed background object at $z$ = 1.847 (see Geach et al.\ 2007). 
It is not clear whether the excess emission 
at 24 $\mu \rm m$ is associated with the lens or the galaxy. However, the 
majority of the radio emission probably originates in the lensing source. The 
steep spectral index calculated between 610 MHz and 1.4 GHz data 
indicates that the 325 MHz flux density should be higher than the catalogue 
entry.

{\bf sxds\_0034}: A compact radio source at $z = 1.095$ with 
a clear QSO component at the wavelengths probed by Spitzer data, a point-like 
optical structure, 
and high-excitation narrow lines in its spectrum. It is, most plausibly, a 
Doppler-boosted object with low extinction and possible time variability. 
This is the one object with significant accretion that lies below the 
FRI/FRII break (Fig.~\ref{ld_sxds}), indicating that it is a FRI `radio-quiet' 
quasar.

{\bf sxds\_0035}: A compact radio source and a galaxy without spectroscopy 
or $J-$/$K-$band data. The HYPERz code fits the 4000$~ \rm \AA$ break 
between $J$ and $K$, which gives a photometric redshift of $z = 2.715$. 

{\bf sxds\_0036}: A compact radio source and a galaxy `G' at $z = 0.871$. The 
610 MHz flux density is an upper limit, giving an uncertain measurement for 
the radio spectral index.

{\bf sxds\_0037}: A compact radio source and a galaxy `G' that is not 
observed in the $J$ and $K$ band (Fig.~\ref{radec}), and which therefore 
has a tentative photometric redshift. This radio source was outside the 
coverage of the 610 MHz survey. We use a typical flat radio spectral 
index $\alpha$ = 0.4, consistent with the compact radio structure of this 
source; we don't trust the 325-MHz catalogue to be complete at this flux 
density level.

\clearpage

\addtocounter{figure}{0}

\begin{figure}
\begin{center}
\setlength{\unitlength}{1mm}
\begin{picture}(50,220)
\put(-10,100){\includegraphics{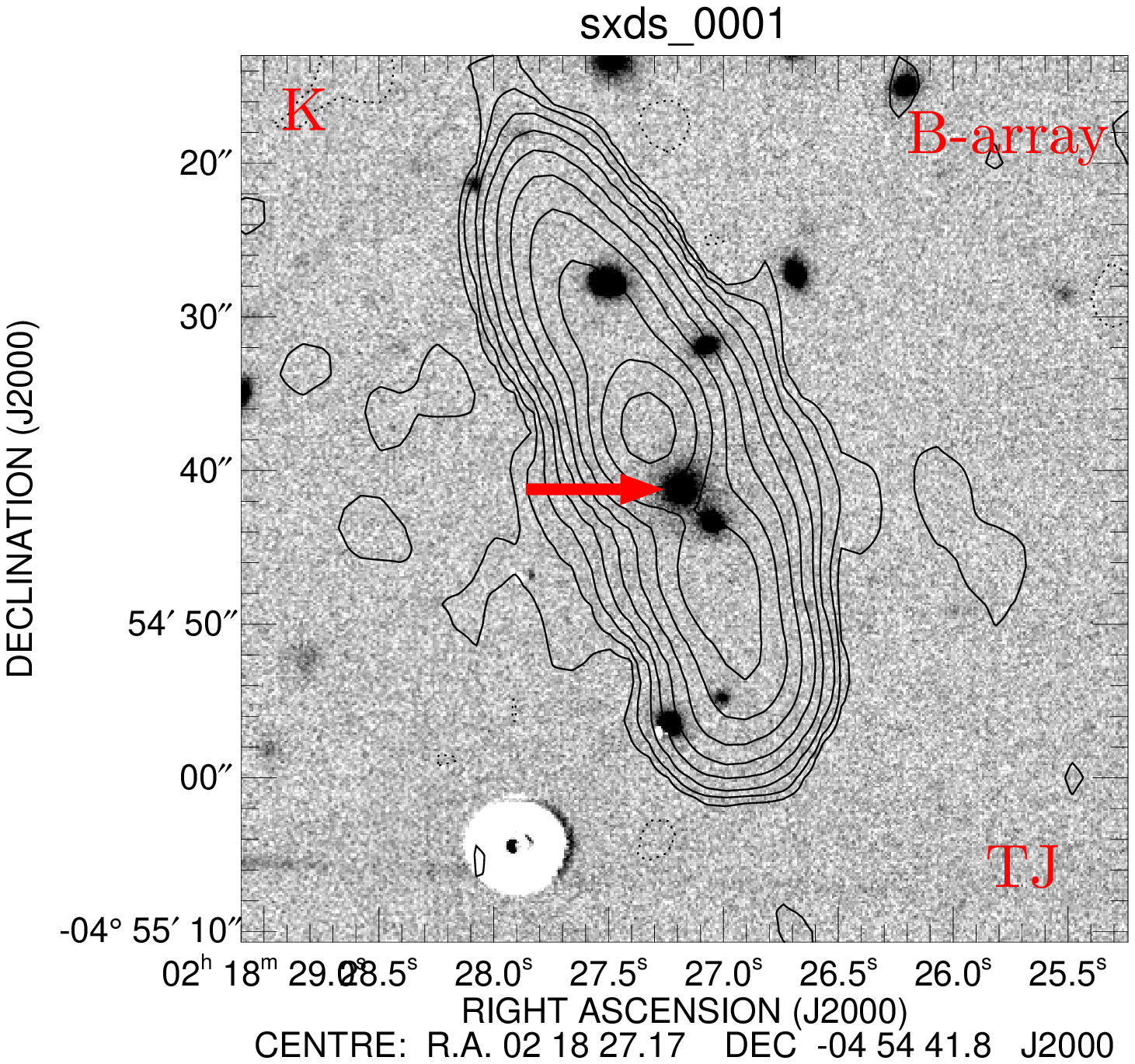}}
\put(12,96){\includegraphics{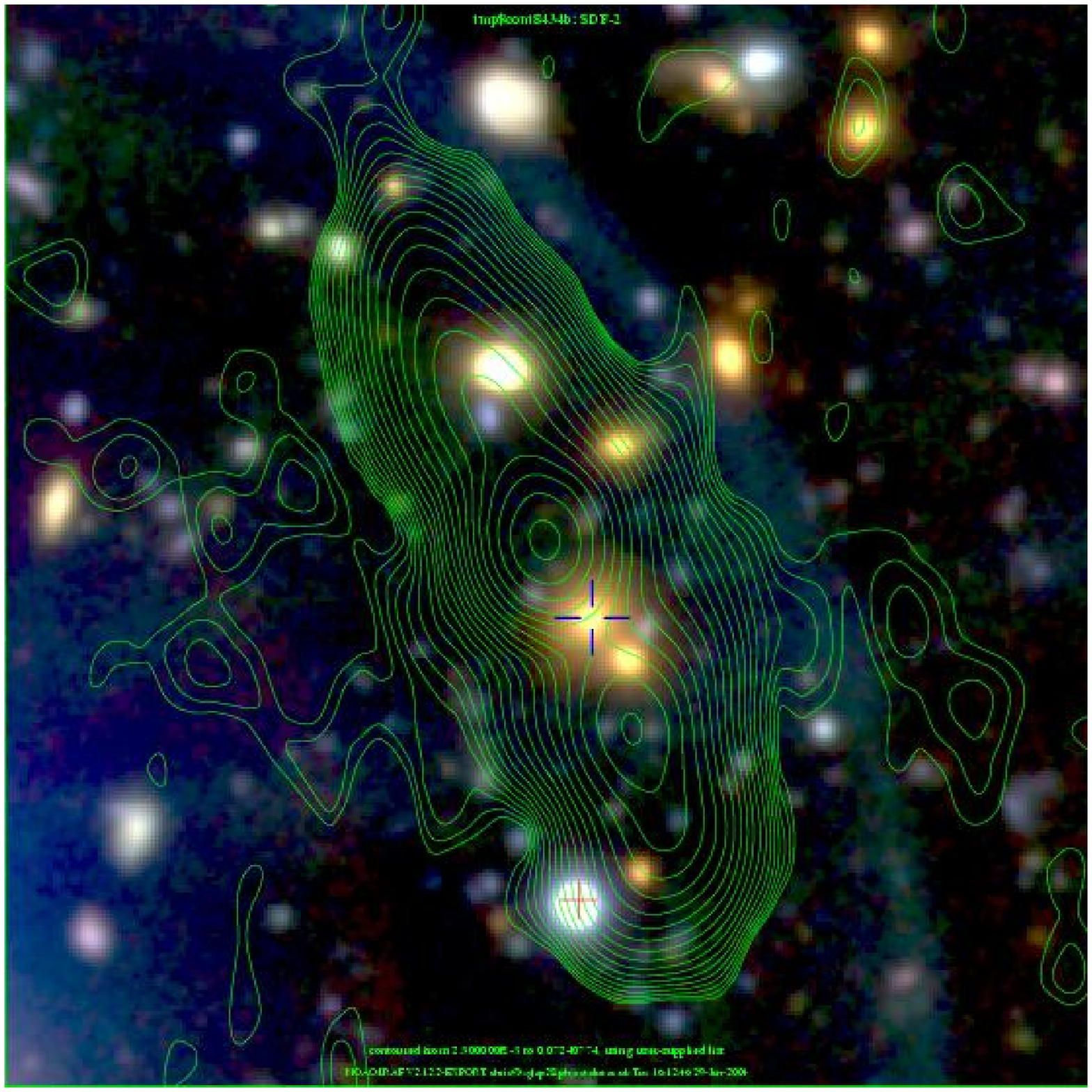}}
\put(57,43){\includegraphics{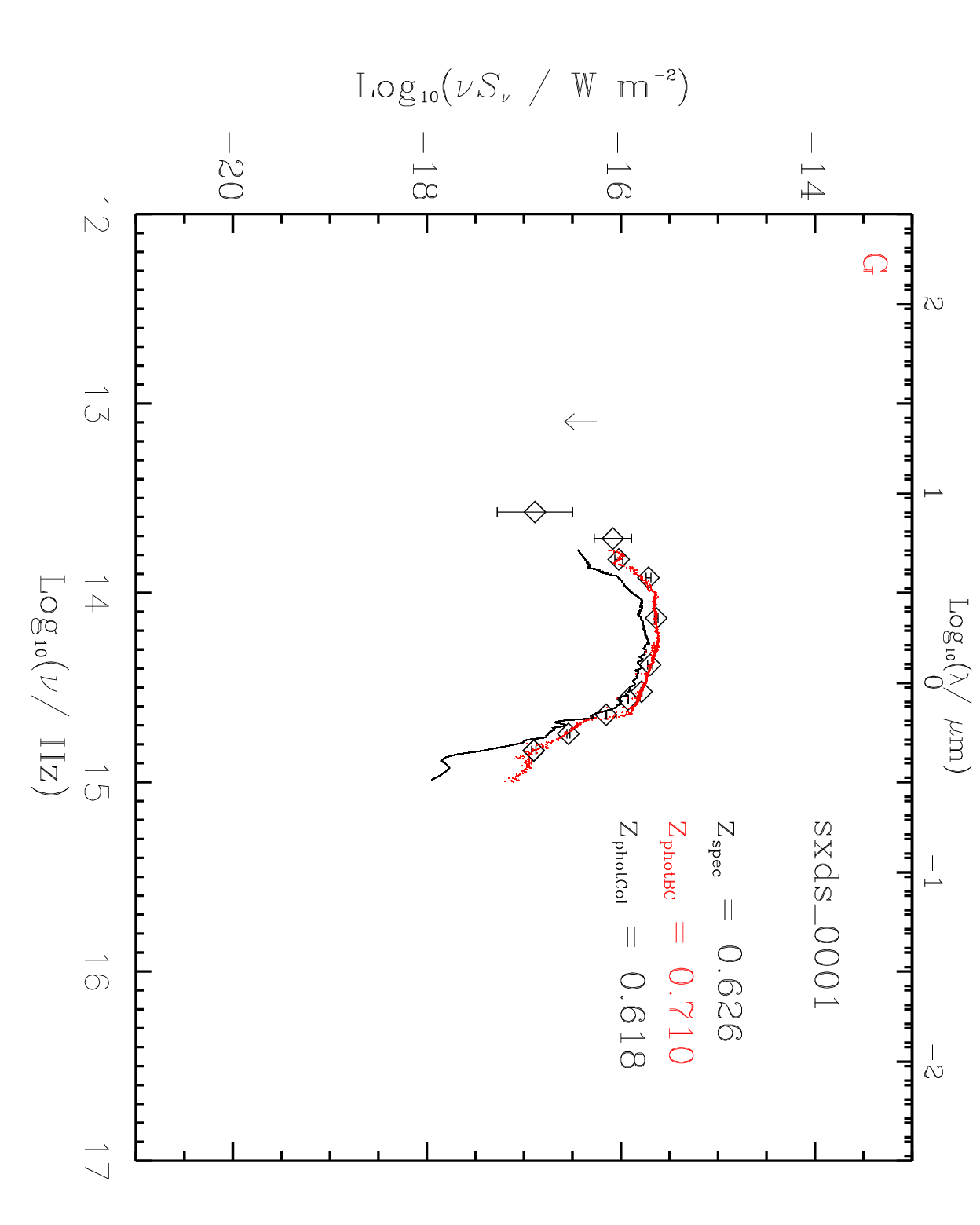}}
\put(90,-26){\includegraphics{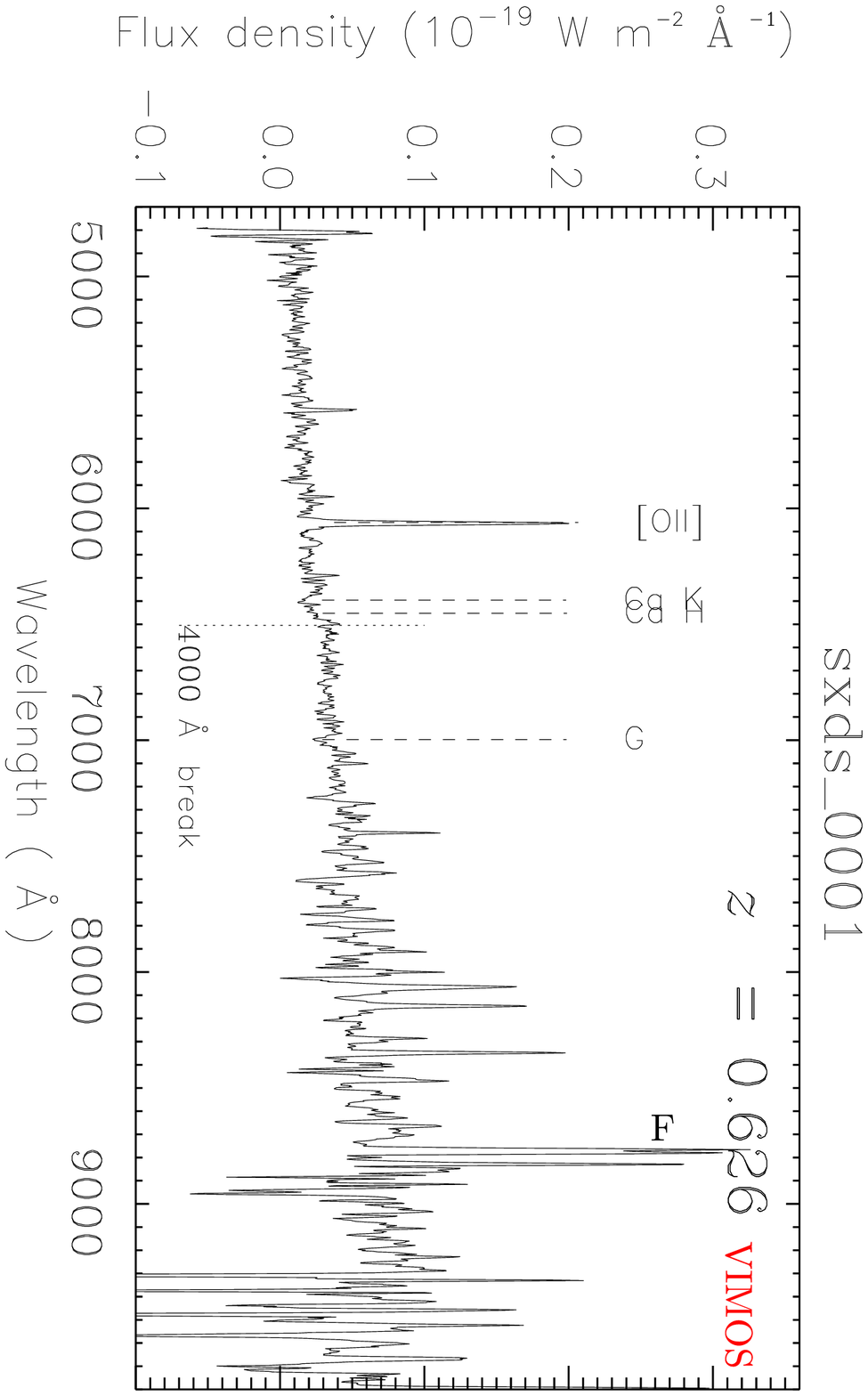}}
\end{picture}
\end{center}

\end{figure}
\clearpage

\addtocounter{figure}{-1}

\begin{figure*}
{\caption[junk]{The rest of Fig. A1 can be found at the URL: 
$www-astro.physics.ox.ac.uk/\sim eleniv/figure\_A1.html$
}}
\begin{center}
\setlength{\unitlength}{1mm}
\begin{picture}(150,220)
\put(-25,-40){\includegraphics{empty.ps}}
\end{picture}
\end{center}
\vspace{0.1in}

\end{figure*}
\addtocounter{figure}{0}


\begin{thebibliography}{}

\bibitem{} Antonucci, R., 1993, ARA\&A, 31, 473
\bibitem{} Aretxaga, I., Hughes, D. H., Coppin, K., Mortier, A. M. J., Wagg, J., Dunlop, J. S., Chapin, E. L., Eales, S. A., Gaztanaga, E., Halpern, M. et al., 2007, MNRAS, 379, 1571
\bibitem{} Bertin, E., Arnouts, S., 1996, AAPS, 117, 393
\bibitem{} Best, P. N., Kauffmann, G., Heckman, T. M., Brinchmann, J., Charlot, S., Ivezic, Z., White, S. D. M., 2005, MNRAS, 362, 25B
\bibitem{} Best, P. N., Arts, J. N., Röttgering, H. J. A., Rengelink, R., Brookes, M. H., Wall, J., 2003, MNRAS, 346, 627
\bibitem{} Blundell, K. M., Rawlings, S., 2001, ApJ, 562, 5
\bibitem{} Blundell, K. M., Rawlings, S., Willott, C. J., 1999, AJ, 117, 677
\bibitem{} Bolzonella, M., Miralles, J. M., Pello, R., 2000, A\&A, 363, 476
\bibitem{} van Breukelen, C., Cotter, G., Rawlings, S., Readhead, T., Bonfield, D., Clewley, L., Ivison, R., Jarvis, M., Simpson, C., Watson, M., 2007, $\rm astro-ph, arXiv:0708.3838$
\bibitem{} van Breukelen, C., Clewley, L., Bonfield, D. G., Rawlings, S., Jarvis, M. J., Barr, J. M., Foucaud, S., Almaini, O., Cirasuolo, M., Dalton, G., Dunlop, J. S., Edge, A. C., Hirst, P., McLure, R. J., Page, M. J., Sekiguchi, K., Simpson, C., Smail, I., Watson, M. G., 2006, MNRAS, 373, L26
\bibitem{} Bruzual, A. G., Charlot, S., 2003, MNRAS, 344, 1000
\bibitem{} Bruzual A. G., Charlot, S., 1993, ApJ, 405, 538
\bibitem{} Cao, X., Rawlings, S., 2004, MNRAS, 349, 1419
\bibitem{} Cimatti, A., di Serego Alighieri, S., Fosbury, R. A. E., 1993, Salvati, M., Taylor, D., 1993, MNRAS, 264, 421
\bibitem{} Coleman, G. D., Wu, C. C., Weedman, D. W., 1980, ApJS , 43, 393
\bibitem{} Condon, J. J., Cotton, W. D., Greisen, E. W., Yin, Q. F., Perley, R. A., Taylor, G. B., Broderick, J. J., 1998, AJ, 115, 1693
\bibitem{} Croton, D. J., Springel, V., White, S. D. M., De Lucia, G., Frenk, C. S., Gao, L., Jenkins, A., Kauffmann, G., Navarro, J. F., Yoshida, N., 2006, MNRAS, 365, 11
\bibitem{} Dunlop, J. S., Peacock, J. A., 1990, MNRAS, 247, 19
\bibitem{} Fanaroff, B. L., Riley, J. M., 1974, MNRAS, 167, 31
\bibitem{} Ferrarese, L., Merritt, D., 2000, ApJ, 539, 9
\bibitem{} Francis, P. J., Hewett, P. C., Foltz, C. B., Chaffee, F. H., Weymann, R. J., Morris, S. L., 1999, ApJ, 373, 465
\bibitem{} Furusawa et al. 2007, accepted by ApJS
\bibitem{} Geach, J. E., Simpson, C. J., Rawlings, S., Read, A. M., Watson, M., 2007, $\rm astro-ph, arXiv:0708.0982$
\bibitem{} Gebhardt, K., Bender, R., Bower, G., Dressler, A., Faber, S. M., Filippenko, A. V., Green, R., Grillmair, C., Ho, L. C., Kormendy, J. et al., 2000, ApJ, 539, 13
\bibitem{} Goto, T., 2007, MNRAS, 381, 187
\bibitem{} Granato, G. L., Danese, L., 1994, MNRAS, 268, 235
\bibitem{} Hambly, N. C., Collins, R. S., Cross, N. J. G., Mann, R. G., Read, M. A., Sutorius, E. T. W., Bond, I. A., Bryant, J., Emerson, J. P., Lawrence, A. et al. 2007, arXiv:0711.3593
\bibitem{} Heywood, I., Blundell, K. M., Rawlings, S., 2007, $\rm astro-ph, arXiv0708.1145$
\bibitem{} Hine, R. G., Longair, M. S., 1979, MNRAS, 188, 111
\bibitem{} Ivison, R. J., Greve, T. R., Dunlop, J. S., Peacock, J. A., Egami, E., Smail, Ian, Ibar, E., van Kampen, E., Aretxaga, I., Babbedge, T., Biggs, A. D., Blain, A. W., Chapman, S. C., Clements, D. L., Coppin, K., Farrah, D., Halpern, M., Hughes, D. H., Jarvis, M. J., Jenness, T., Jones, J. R., Mortier, A. M. J., Oliver, S., Papovich, C., Pérez-González, P. G., Pope, A., Rawlings, S., Rieke, G. H., Rowan-Robinson, M., Savage, R. S., Scott, D., Seigar, M., Serjeant, S., Simpson, C., Stevens, J. A., Vaccari, M., Wagg, J., Willott, C. J., 2007, MNRAS, 380, 199
\bibitem{} Jackson, C. A., Wall, J. V., 2001, ASPC, 227, 242
\bibitem{} Jannuzi, B., Elston, R., Schmidt, G. D., Smith, P. S., Stockman, H. S., 1995, AJ, 454, 111
\bibitem{} Johnston, H., Hunstead, R., Sadler, E., Cotter, G., Morganti, R., 2004, IAUS, 222, 451
\bibitem{} Kennicutt, R. C., 1998, ARA\&A, 36, 189
\bibitem{} Lavalley, M., Isobel, T., Feigelson, E., 1992, ASPC, 25, 245
\bibitem{} Lawrence, A., Warren, S. J., Almaini, O., Edge, A. C., Hambly, N. C., Jameson, R. F., Lucas, P., Casali, M., Adamson, A., Dye, S., et al., 2007, MNRAS, 379, 1599
\bibitem{} Ledlow, M. J., Owen, F. N., 1996, IAUS, 175, 238
\bibitem{} Liu, F. K., Wu, X-B., Cao, S. L., 2003, MNRAS, 340, 411
\bibitem{} Lonsdale, C. J., Smith, H. E., Rowan-Robinson, M., Surace, J., Shupe, D., Xu, C., Oliver, S., Padgett, D., Fang, F., Conrow, T. et al., 2003, PASP, 115, 897
\bibitem{} Magorrian, J., 2006, MNRAS, 373, 425
\bibitem{} Martinez-Sansigre, A., Rawlings, S., Garn, T., Green, D. A., Alexander, P., Klöckner, H. R., Riley, J. M., 2006, MNRAS, 373, 80
\bibitem{} McLure, R. J., Dunlop J. S., 2004, MNRAS, 352, 1390
\bibitem{} McLure, R. J., Willott, C. J., Jarvis, M. J., Rawlings, S., Hill, G. J., Mitchell, E., Dunlop, J. S., Wold, M. , 2004, MNRAS, 351, 347
\bibitem{} Mortier, A. M. J., Serjeant, S., Dunlop, J. S., Scott, S. E., Ade, P., Alexander, D., Almaini, O., Aretxaga, I., Baugh, C., Benson, A. J., et al., 2005, MNRAS, 363, 563
\bibitem{} Ogle, P., Whysong, D., Antonucci, R., 2006, ApJ, 647, 161
\bibitem{} Owen, F. N., Laing, R. A., 1989, MNRAS, 238, 357
\bibitem{} Pierre, M., Valtchanov, I., Alrieri, B., et al., 2004, J. Cosmol. Astro-Part. Phys., 9, 11
\bibitem{} Rawlings, S., Saunders, R., 1991, Nature, 349, 138
\bibitem{} Richards, G. T., Hall, P. B., Vanden Berk, D. E., Strauss, M. A., Schneider, D. P., Weinstein, M. A., Reichard, T. A., York, D. G., Knapp, G. R., Fan, X., Ivezic, Z., Brinkmann, J., Budavari, T., Csabai, I., Nichol, R. C., 2003, AJ, 126, 1131
\bibitem{} Rowan-Robinson, M., 1995, MNRAS, 272, 737
\bibitem{} Sarazin, C. L., Koekemoer, A. M., Baum, S. A., O'Dea, C. P., Owen, F. N., Wise, M. W., 1999, ApJ, 510, 90
\bibitem{} Schoenmakers, A. P., de Bruyn, A.G., Röttgering, H.J.A., van der Laan, H., Mack, K.-H., Kaiser, C.R.. 1999, $\rm astro-ph/9910448$
\bibitem{} Sekiguchi, K., 2004, AAS, 205, 8105
\bibitem{} di Serego Alighieri, S., Cimatti, A., Fosbury, R. A. E., 1993, ApJ, 404, 584
\bibitem{} Serjeant, S., Rawlings, S., Lacy, M., McMahon, R, G., Lawrence, A., Rowan-Robinson, M., Mountain, M., 1998, MNRAS, 298, 321
\bibitem{} Simpson, C., Martinez-Sansigre, A., Rawlings, S., Ivison, R., Akiyama, M., Sekiguchi, K., Takata, T., Ueda, Y., Watson, M., 2006, MNRAS, 372, 741
\bibitem{} Tadhunter, C. N., Scarrott, S. M., Draper, P., Rolph, C., 1992, MNRAS, 256, 53
\bibitem{} Tasse, C., Cohen, A. S., Rottgering, H. J. A., Kassim, N. E., Pierre, M., Perley, R., Best, P., Birkinshaw, M., Bremer, M., Liang, H., 2006, A\&A, 456, 791
\bibitem{} Vardoulaki, E., Rawlings, S., Simpson, C., 2006, $\rm astro-ph/0609719$
\bibitem{} Vardoulaki, E., Rawlings, S., Hill, G. J., Croft, S., Brand, B., Riley, J., Willott, C., 2006, AN, 327, 282
\bibitem{} Willott, C. J., Rawlings, S., Blundell, K. M., Lacy, M., 2000, MNRAS, 316, 449
\bibitem{} Willott, C. J., Rawlings, S., Jarvis, M. J., Blundell, K. M., 2003, MNRAS, 339, 173



\end{thebibliography}
\end{document}